\documentclass[aps,twocolumn,preprintnumbers,amsmath,amssymb]{revtex4}

\usepackage{epsfig}
\usepackage{epstopdf}
\usepackage{graphicx}
\usepackage{dcolumn}
\usepackage{bm}
\usepackage{subfigure}
\usepackage{booktabs}

\begin{document}

\title[Probing the anisotropic local universe with SNe\,Ia]{Probing
  bulk flow with nearby SNe\,Ia data}
\date{\today}

\author{${}^{a}$Stephen Appleby, ${}^{b}$Arman Shafieloo, ${}^{c,d}$Andrew Johnson}
\affiliation{${}^{a}$Asia Pacific Center for Theoretical Physics, Pohang, Gyeongbuk 790-784, Korea \\
${}^{b}$Korea Astronomy and Space Science Institute, Daejeon, 305-348, Korea\\ 
${}^c$Centre for Astrophysics \& Supercomputing, Swinburne University of Technology, P.O. Box 218, Hawthorn, VIC 3122, Australia.\\ 
${}^d$ARC Centre of Excellence for All-sky Astrophysics (CAASTRO)\\}

\date{\today}

\begin{abstract}
We test the isotropy of the local Universe using low redshift supernova data from various catalogs and the non-parametric method of smoothed residuals. Using a recently developed catalog which combines supernova data from various surveys, we show that the isotropic hypothesis of a Universe with zero velocity perturbation can be rejected with moderate significance, with $p$-value $\sim 0.07$ out to redshift $z < 0.045$. We estimate the direction of maximal anisotropy on the sky for various pre-existing catalogs and show that it remains relatively unaffected by the light curve fitting procedure. However the recovered direction is biased by the underlying distribution of data points on the sky. We estimate both the uncertainty and bias in the direction by creating mock data containing a randomly oriented bulk flow and using our method to reconstruct its direction. We conclude that the inhomogeneous nature of the data introduces a directional bias in galactic latitude of approximately $|\Delta b_{\rm max}| \sim 18^{\circ}$ for the supernova catalog considered in this work, and after correcting for this effect we infer the direction of maximum anisotropy as $(b,\ell) = (20^{\circ}, 276^{\circ}) \pm (12^{\circ},29^{\circ})$ in galactic coordinates. Finally we compare the anisotropic signal in the data to mock realisations in which large scale velocity perturbations are consistently accounted for at the level of linear perturbation theory. We show that including the effect of the velocity perturbation in our mock catalogs degrades the significance of the anisotropy considerably, with $p$-value increasing to $\sim 0.29$. One can conclude from our analysis that there is a moderate deviation from isotropy in the supernova data, but the signal is consistent with a large scale bulk velocity expected within $\Lambda$CDM. 
\end{abstract}

\maketitle 

\section{Introduction}
\label{sec:introduction}

The use of type Ia supernovae (SN) as a tool to probe cosmology has a long and distinguished history, starting with early pioneering works \cite{Riess:1998cb,Perlmutter:1998np,Garnavich:1998th,Riess:1998dv} and continuing to the present \cite{Blakeslee:2003dz,Tonry:2003zg,Riess:2003gz,Barris:2003dq,Matheson:2004dj,Hicken:2009dk,Hicken:2009df,Hamuy:2005tf,Folatelli:2009nm,Astier:2005qq, Sako:2007ms,Miknaitis:2007jd,WoodVasey:2007jb,Holtzman:2008zz,Kessler:2009ys,Freedman:2009vv}. Various existing catalogs \cite{Tonry:2003zg,Ganeshalingam:2013mia,Amanullah:2010vv,Hicken:2009dk,Kessler:2009ys,Folatelli:2009nm}, containing hundreds of the standardizeable candles, have been used to infer the existence of a late time accelerating epoch with strong statistical significance \cite{WoodVasey:2007jb,Suzuki:2011hu,Sullivan:2011kv}. Their complementarity to other cosmological data sets \cite{Tegmark:1998yy} allows one to set stringent constraints on the now widely accepted $\Lambda$CDM model. 
       
To maximize the constraining power of supernovae, one must use a combination of both high and low redshift data. The low redshift supernova play an important role, acting essentially as an anchor in the redshift luminosity distance relation \cite{Hamuy:2005tf}. They are relatively insensitive to cosmological parameters, however their worth lies in allowing us to determine the relative brightness of the supernova. Unfortunately existing low redshift samples are sensitive to comparatively large systematic uncertainties, in particular calibration issues and host mass - SN brightness correlations. Future surveys are expected to provide a more homogeneous set of accurately calibrated objects. When using current data however, it is important that we understand and account for the various measurement and astrophysical uncertainties.

One phenomenon that will affect the low redshift data is the existence of peculiar velocities. It is well known that the local group has a velocity $V_{\rm bulk} = 627 \pm 22 {\rm km s^{-1}}$ relative to the CMB rest frame, in the direction $(b,\ell) \sim (30^{\circ},276^{\circ})$ \cite{Kogut:1993ag}. The origin of this motion is typically attributed to low redshift superclusters, whose gravitational attraction will produce coherent velocities in nearby galaxies. However, recent claims of a detection of a coherent bulk flow out to very large distances $d_{\rm c} \sim 800 {\rm Mpc}$ \cite{Kashlinsky:2008ut,Kashlinsky:2009dw,Kashlinsky:2008us,Kashlinsky:2010ur} have led to more exotic explanations such as an inhomogeneous pre-inflationary spacetime \cite{Turner:1991dn} or higher dimensional gravity theories \cite{Afshordi:2008rd}. Large scale cosmological anisotropies can also be generated by magnetic fields produced by Lorentz violating effects during inflation \cite{Campanelli:2009tk}.

Assuming that the bulk flow is due to local attractors, the lack of all sky and deep peculiar velocity data means that one cannot yet definitively answer the question as to what mass distributions are responsible for the observed flow. Nor can we deduce at what scale the coherent motion ceases to be significant. The issue has been considered for decades, and known superclusters such as Shapley and the so-called {\it Great Attractor} are commonly thought to contribute at least partially to the observed bulk motion \cite{LyndenBell:1988qs,Scaramella89}. More recently, different groups have used various data sets \cite{Watkins:2008hf,Turnbull:2011ty,Colin:2010ds,Nusser:2011sd,Nusser:2011tu,Branchini:2012rb,Ma:2012tt} to estimate the direction and magnitude of the flow at different scales. While consensus appears to have been reached regarding the direction, the magnitude and consistency with $\Lambda$CDM remains an open question. It is an issue that has drawn scrutiny from various sources, and remains contested \cite{Keisler:2009nw,Osborne:2010mf,Mody:2012rh,Lavaux:2012jb,Kashlinsky:2008ut}. Some groups have suggested that supernova at redshift $z \gtrsim 0.05$ could exhibit evidence of back-infall into the Shapley supercluster, indicating that it might be the source of the observed bulk motion \cite{Colin:2010ds}.

Supernova are ideal candidates to study the local bulk flow, being bright distance indicators with  precise redshift measurements. The depth of supernova surveys is typically large, allowing us to probe the scale at which the bulk flow remains coherent. However this gain in depth relative to local galaxy catalogs is offset by the comparatively modest quantity of existing data. Future surveys such as the Large Synaptic Survey Telescope \cite{Abate:2012za} are expected to yield thousands of supernova and should definitively answer the question as to the magnitude and direction of the bulk flow. The increasing number of detected SN will allow us to move beyond simply treating the peculiar velocities as systematics in distance measurements. The supernovae measurements yield information regarding the peculiar velocity field, and open up a new avenue in exploring this dynamical cosmological probe \cite{Johnson:2014kaa}.

The question that we address in this work is whether one can observe a statistically significant deviation from isotropy in existing low redshift supernova data, and if so whether this anisotropic signal is consistent with the standard cosmological model. To achieve this aim we adopt a non-parametric method, which compares the distance modulus residuals to mock realisations drawn from the underlying covariance matrix of the data \cite{Colin:2010ds}. In this way we address the significance of any anisotropic signal within the context of the systematic and statistical uncertainties associated with the data. We make no assumptions regarding the nature of any potential anisotropy. This is an advantage in the sense that we do not need to specify a model and can detect any source of anisotropic behaviour. However a consequence of this model independence is that one cannot measure the magnitude of the bulk flow using our approach.

The paper will proceed as follows. In the following section \ref{sec:2} we briefly review the method used to test the isotropic hypothesis. We discuss the catalogs used and how we combine them in section \ref{sec:3}, and present our results in \ref{sec:4}. We discuss the effect of inhomogeneous data distributions and the effect of correlations, and conclude in \ref{sec:7}.

\section{\label{sec:2}The method of residuals}

The method of residuals has been discussed in detail in \cite{Colin:2010ds,Feindt:2013pma,Appleby:2013ida}, and we begin with a brief review. We note that numerous other approaches to testing isotropy exist in the literature, and we direct the reader to \cite{Kolatt:2000yg,Bonvin:2006en,Gordon:2007zw,Schwarz:2007wf,Gupta:2007pb,Appleby:2012as,Cooray:2008qn,Gupta:2010jp, Cooke:2009ws,Koivisto:2008ig,Koivisto:2010dr,Campanelli:2007qn,Campanelli:2006vb,Appleby:2009za, Antoniou:2010gw,Blomqvist:2008ud,Blomqvist:2010ky,Tsagas:2009nh,Appleby:2014lra} for searches of (predominantly cosmological) anisotropic signals. The effect of peculiar velocities on supernovae data has been discussed in \cite{Hui:2005nm,Davis:2010jq}.

Our approach entails creating a map on the sky of the sum of distance modulus residuals, smoothed over a particular scale $\delta$. The first step is to calculate a global best fit cosmology. We fix the background expansion to be flat, $\Lambda$CDM and fit the parameter $\Omega_{\rm m 0}$ to our chosen SN catalog by minimizing the $\chi^{2}$ distribution

\begin{equation} \chi^{2} =  \delta {\bf \mu}^{\rm T} \Sigma^{-1} \delta {\bf \mu}  \end{equation}

\noindent where $\delta {\bf \mu}$ is the data vector of distance modulus residuals - $\delta \mu_{\rm i} = \mu_i(z_i)  - \mu_{\rm th}(z_i)$, and $\Sigma$ is the covariance matrix of the data. $\mu_{\rm i}(z_{\rm i})$ is the distance modulus of the $i^{\rm th}$ data point and $\mu_{\rm th}(z_{\rm i})$ is the theoretically predicted $\Lambda$CDM value at data point redshift $z_{\rm i}$. Our approach to calculating the best fit cosmology differs from more detailed cosmological parameter estimation considered in \cite{Kowalski:2008ez,Amanullah:2010vv,Suzuki:2011hu}, for example. In the Union 2.1 catalog the intrinsic dispersion $\sigma_{\rm i}$ is kept as a free parameter to account for the intrinsic scatter of the supernova magnitudes, and the cosmology fit such that $\chi^{2} = 1$ (per degree of freedom). However, the best fit cosmological parameters that we obtain are essentially identical to the values quoted in the various catalogs that we will use. Furthermore, the cosmological dependence of our result is expected to be very weak, as we are focusing solely on the low redshift $z < 0.1$ SNe sample. 

We denote the best-fit distance modulus as $\tilde\mu(z,H_{0},\Omega_{\rm m0})$. Using this, we construct the error-normalised residuals of the data from the model \cite{Perivolaropoulos:2008yc},

\begin{equation} \label{eq:res1}
  q_i (z_i, \theta_i, \phi_i) = \frac{\mu_i(z_i, \theta_i, \phi_i) 
    - \tilde\mu_i(z_i)}{\sigma_i(z_i)}\;, 
\end{equation}

\noindent Here, $(\theta_i, \phi_i)$ are the angular positions of the $i^{\rm th}$ data point on the sphere and $\sigma_{\rm i}(z_{\rm i})$ is the error associated with the diagonal component of the covariance matrix. 

Next, we define a measure $Q (\theta, \phi)$ on the surface of the
sphere using these residuals
\begin{equation} \label{eq:Q}
  Q (\theta,\phi) = \sum_{i=1}^{N_{\rm SN}} q_i (z_i, \theta_i, \phi_i)
   W (\theta, \phi, \theta_i, \phi_i)\;, 
\end{equation} 

\noindent where $N_{\rm SN}$ is the number of SNe\,Ia data points and $W (\theta, \phi,
\theta_i,\phi_i)$ is a weight (or window) function that represents a two dimensional
smoothing. We define the weight using a Gaussian distribution

\begin{equation}
  W (\theta, \phi, \theta_i, \phi_i) = \frac{1}{\sqrt{2\pi}\delta} 
   \exp\left[-\frac{L(\theta, \phi, \theta_i, \phi_i)^2}{2\delta^2}\right]\;,
\end{equation} 

\noindent where $\delta$ is the width of smoothing and $L (\theta, \phi,
\theta_i, \phi_i)$ is the distance on the surface of a sphere of unit
radius between two points with spherical coordinates $(\theta, \phi)$
and $(\theta_i, \phi_i)$

\begin{equation}
  L (\theta, \phi, \theta_i, \phi_i) = 2 \arcsin \frac{R}{2}\;, 
\end{equation} 

\noindent where 

\begin{eqnarray*}
R=\Big(\big[\cos(\theta_i)\cos(\phi_i)-\cos(\theta)\cos(\phi) \big]^2 +\\
\big[\cos(\theta_i)\sin(\phi_i)-\cos(\theta)\sin(\phi) \big]^2 
+ \big[\sin(\theta_i)-\sin(\theta)\big]^2 \Big) ^{1/2}.
\end{eqnarray*}

\noindent Any anisotropy in the data will translate to $Q (\theta, \phi)$ deviating from zero.  
Finally we adopt a value for $\delta$, calculate $Q (\theta, \phi)$ on the
whole surface of the sphere and find the extreme values of this function. In this work we utilize the maximum dipole,  

\begin{equation} 
  \Delta Q_{\rm d}   = Q (\theta, \phi) 
   - Q (\theta_{\rm d}, \phi_{\rm d}) ,
\label{DeltaQ}
\end{equation}

\noindent where $(\theta_{\rm d},\phi_{\rm d})$ are the angles that locate the dipole of $(\theta,\phi)$ on the unit sphere. 
In what follows we choose the galactic coordinate system $(\theta,\phi) = (b, \ell)$. In this case, we have $(\theta_{\rm d},\phi_{\rm d})$ and $(-b,\ell+180^{\circ})$ if $\ell<180^{\circ}$ and $(-b,\ell-180^{\circ})$ for $\ell>180^{\circ}$. In principle one could use any number of different measures of anisotropy - for example the maximum and minimum of this function on the sky $Q_{\rm max}$ and $Q_{\rm min}$. However for practical purposes current supernova data is neither sufficiently copious or precise to detect anything other than large scale flows such as the dipole. Hence our use of $\Delta Q_{\rm d}$.

A large value of $\Delta Q_{\rm d}$ implies a significant anisotropy in the data. However we expect this quantity, which denotes the extreme of the $Q(\theta,\phi)$ function on the sphere, will not be zero even if the Hubble residuals are isotropic on the sky. We must test the significance of the magnitude of $\Delta Q_{\rm d}$ by creating mock realisations.

In this work we construct two sets of $N_{\rm real} = 1000$ realisations of our SN catalogs. We describe both in turn below, and in what follows will refer to them as realisations A and B respectively. 

In the first case we simply follow the original method of \cite{Colin:2010ds} and construct our realisations by holding each data point at fixed redshift and position on the sky, and estimate the distance modulus as the isotropic best fit value $\tilde{\mu}(z_{\rm i},\Omega_{\rm m0})$ plus a contribution drawn from a Gaussian of width $\sigma_{\rm i}$, where $\sigma_{\rm i}$ is the observational uncertainty on the $i^{\rm th}$ data point quoted in the catalog. Hence we have $\mu_{\rm i} = \tilde{\mu}(z_{\rm i},\Omega_{\rm m0})+\delta \mu_{\rm i, G}$, where $\delta \mu_{\rm i, G}$ is the Gaussian component. For each data realisation, $\Delta Q_{\rm d}$ is obtained and an empirical Probability Distribution Function (PDF) is constructed for this function. We then ascertain the significance of $\Delta Q_{\rm d}$ obtained from the data by calculating its p-value from the PDF. When comparing the data to this set of realisations, we are addressing the question as to whether any potential anisotropy detected in the data is consistent with observational uncertainties within the data, which we are taking to be Gaussian and uncorrelated. Note that we are implicitly assuming that the error is Gaussian here - for this reason our choice of observable is the distance modulus rather than the velocity \cite{Strauss:1995fz}. The p-value in this case signifies the frequency of occurrence of a $\Delta Q_{\rm d}$ value obtained from the simulations that is greater than or equal to the data value of $\Delta Q_{\rm d}$. The simulations are drawn from an isotropic $\Lambda$CDM model with no anisotropic signal, so one can interpret the $p$-value as the probability that the data value of $\Delta Q_{\rm d}$ can be obtained in an isotropic Universe.

Our primary aim is to test the consistency of the data with the underlying cosmological model. Inhomogeneous structure present in the late time Universe generates coherent velocity flows, which breaks isotropy at low redshift. One can estimate the expected magnitude of the velocity field on large scales using linear perturbation theory. For the second set of simulated realisations, we use the covariance matrix constructed in \cite{Johnson:2014kaa} which accounts for both observational uncertainties and the presence of coherent large scale peculiar velocities. The catalog is discussed further in section \ref{sec:3}, here we briefly review the contributions to the covariance matrix. 

The covariance between the $i^{\rm th}$ and $j^{\rm th}$ data point - $C_{ij}$ - is given by \cite{Hui:2005nm,Johnson:2014kaa}

\begin{widetext}

\begin{equation}\label{eq:cov}  C_{ij} = \left({5 \over \ln 10} \right)^{2} \left( 1 - {(1+z_{i})^{2} \over H(z_{i})d_{\rm L}(z_{i})}\right) \left( 1 - {(1+z_{j})^{2} \over H(z_{j})d_{\rm L}(z_{j})}\right) \int {dk \over 2\pi^{2}} k^{2} P_{\rm vv}(k,a=1)W(k,\alpha_{ij},r_{i},r_{j}) \end{equation}

\end{widetext}

\noindent where $z_{i}$, $z_{j}$ are the redshifts of the data points, and $H(z_{i,j})$ $d_{\rm L}(z_{i,j})$ the corresponding Hubble parameter and luminosity distance at said redshifts. $P_{\rm vv}(k,a=1)$ is the linear velocity power spectrum obtained for our best fit cosmological parameters, evaluated at $z=0$. The kernel $W(k,\alpha_{ij},r_{i},r_{j})$ is the angular component of the integral over the Fourier modes, and is given by 

\begin{eqnarray} \nonumber & &  W(k,\alpha_{ij},r_{i},r_{j}) = {1 \over 3} \left( j_{0}(kA_{ij}) - 2 j_{2}(kA_{ij} \right) \hat{r}_{i} . \hat{r}_{j} \\
& & \qquad + {1 \over A_{ij}^{2}} j_{2}(kA_{ij}) r_{i}r_{j} \sin^{2}(\alpha_{ij}) \end{eqnarray} 

\noindent where $\alpha_{ij} = \cos^{-1}(\hat{r}_{i} . \hat{r}_{j})$, $A_{ij} = |r_{i} - r_{j}|$, $r_{i}$ is the position vector of the $i^{\rm th}$ data point and $j_{0,2}$ are Bessel functions. See (for example) appendix A of \cite{Ma:2010ps} for a derivation of this function.

The first two redshift dependent cofactors in ($\ref{eq:cov}$) transform from correlations between peculiar velocities to correlations between distance modulus fluctuations \cite{Hui:2005nm}. The integral over the Fourier modes is related to the correlation between data points $i$ and $j$ due to the fact that they trace the same underlying density field - the trace components of $C_{ij}$ are cosmic variance and the off-diagonal terms the cross correlation. The velocity power spectrum $P_{\rm vv}$ is obtained from linear perturbation theory assuming $\Lambda$CDM. Bulk velocities on the scales that are being considered in this work are well modeled using linear theory, and so this contribution to the covariance matrix accounts for large scale velocity components expected within the standard cosmological model. 

To this covariance matrix, the {\it diagonal} component of the catalog observational uncertainty is added according to 

\begin{equation} \label{eq:ctot} \Sigma_{ij} = C_{ij} + \sigma_{i}^{2} \delta_{ij} \end{equation}

\noindent where 

\begin{equation} \sigma_{i}^{2} = \sigma_{\rm obs}^{2} + \left({5 \over \ln 10} \right)^{2} \left( 1 - {(1+z_{i})^{2} \over H(z_{i}) d_{\rm L}(z_{i})}\right)^{2} \sigma_{\rm v}^{2} , \end{equation}

\noindent $\sigma_{\rm obs}$ is the observational error quoted in the catalog and $\sigma_{\rm v}$ is the uncertainty associated with non-linear peculiar velocities of the galaxies. In ref.\cite{Johnson:2014kaa} $\sigma_{\rm v}$ was treated as a free parameter and fit according to the data. Here we fix its value to $\sigma_{\rm v}=400 {\rm km s^{-1}}$, in accordance with \cite{Johnson:2014kaa}.  

To obtain the $N_{\rm real} = 1000$ mock realisations, we take the full covariance matrix ($\ref{eq:ctot}$) and diagonalise it by calculating its Eigenvectors,

\begin{equation} {\bf \Phi^{\rm T} \Sigma \Phi} = {\bf \Lambda } \end{equation} 

\noindent Where ${\bf \Phi}$ is the $N_{\rm SN} \times N_{\rm SN}$ matrix constructed from the Eigenvectors of the full covariance matrix ${\bf \Sigma}$ and ${\bf \Lambda}$ is the diagonal matrix constructed from the Eigenvalues of ${\bf \Sigma}$. If ${\bf \Sigma}$ is the covariance matrix of the distance modulus residual vector $\delta \underline{\mu}$, then ${\bf \Lambda}$ is the covariance matrix of the de-correlated data vector $\delta \underline{\mu}^{\rm (diag)} = {\bf \Phi^{\rm T}} \delta \underline{\mu}$. To incorporate the correlations into our simulations, we obtain our $N_{\rm real}=1000$ realisations by drawing Gaussian realisations from the diagonal covariance matrix ${\bf \Lambda}$ to obtain a mock data vector $\delta \underline{\mu}^{\rm (diag)}$, and then transforming back into the original basis according to $\delta \underline{\mu} = {\bf \Phi} \delta\underline{\mu}^{\rm (diag)}$. These $\delta \mu$ residuals are used to construct a PDF of $\Delta Q_{\rm d}$ in our second set of realisations. Once again, the $p$-value is defined as the fraction of $\Delta Q_{\rm d}$ values from the simulations that are greater than or equal to the data value. The simulations now contain the effect of large scale velocity perturbations on the data, and hence account for correlations between data points due to the fact that they are tracing the same underlying density and velocity fields.

As a final comment in this section, one should be careful when attempting to infer the direction of maximal anisotropy using this method. It is known \cite{Appleby:2013ida} that the method is capable of approximately reconstructing the direction of an anisotropic signal for a homogeneous distribution of data, but the result can be skewed when the data is inhomogeneously scattered on the sky \cite{Feindt:2013pma}. We test the reliability of the method in selecting the `true' anisotropic direction $(b, \ell)$ in section \ref{sec:4}.

\section{\label{sec:3}Supernova catalogs}

We apply the method outlined in section \ref{sec:2} to a number of existing data sets in the literature, specifically the Union 2.1 \cite{Kowalski:2008ez,Amanullah:2010vv,Suzuki:2011hu}, Constitution \cite{Hicken:2009dk} and LOSS \cite{Ganeshalingam:2013mia} samples. Our aim in comparing these different catalogs is to test for consistency amongst the different light curve fitting procedures, to ensure that any anisotropic signal is not due to unknown systematics associated with modeling the data. However the main purpose of this work is to test the consistency of the supernova data with the $\Lambda$CDM model - for this purpose we utilize a recently developed catalog which combines the majority of low redshift SN from a variety of surveys \cite{Johnson:2014kaa}. 

The Union 2.1 sample contains $N_{\rm SN} =175$ supernova out to redshift $z < 0.1$, which is the limit to which we perform our analysis. Beyond this, any effect due to bulk flow velocities will be practically negligible. The data set is an amalgamation from numerous sources - and we direct the reader to \cite{Kowalski:2008ez,Amanullah:2010vv,Suzuki:2011hu} for details. As the SN data arise from a number of different telescopes, each with their own systematics and calibrations, each survey's contribution to the overall set is carefully modified to include effects such as photometric zero point offsets, contamination, Malmquist bias, K-corrections and gravitational lensing. The light curves of the entire sample are analysed using the ${\rm SALT}2$ light curve fitting procedure, in which each supernova are assigned three parameters - the peak magnitude $m_{\rm B}^{\rm max}$, the light curve width $x_{1}$ and $c$, which encodes the effects due to intrinsic colour and dust reddening. The distance modulus is then constructed as 

\begin{equation} \mu_{\rm B} = m_{\rm B}^{\rm max} + \alpha x_{1} - \beta . c + \delta . P(m_{\ast}^{\rm true} < m_{\ast}^{\rm threshold}) - M_{\rm B} \end{equation} 

\noindent where $M_{\rm B}$ is the absolute B-band magnitude of a type Ia supernova with $x_{\rm 1} = 0$, $c=0$ and $P(m_{\ast}^{\rm true} < m_{\ast}^{\rm threshold})=0$. Here $P(m_{\ast}^{\rm true} < m_{\ast}^{\rm threshold})=0$ denotes the host mass correction that accounts for the correlation between the luminosity of the SN and the host galaxies mass \cite{Suzuki:2011hu}.  

The Constitution catalog contains $N_{\rm Cons}=256$ supernovae, also obtained from a variety of surveys. It shares many of the same low redshift supernova as Union 2.1, however the method of dealing with systematics and choice of light curve fitting procedure vary. The dominant systematics are identified as the choice of nearby training set used, and the light curve fitting procedure adopted. Specifically, the treatment of host galaxy reddening is the primary source of systematic uncertainty. To examine these effects four different light curve fitting procedures are utilised in \cite{Hicken:2009dk}. SALT2 is used in the same manner as in the Union $2.1$ sample - where all host reddening effects are incorporated via the color term $c$, with the empirical relation $\mu_{\rm B} \propto c$. 

The MLCS light curve fitting method utilises a different approach - fitting the SN distance in conjunction with a shape/luminosity parameter $\Delta$ and the host galaxy extinction parameter $A_{\rm V}$. The extinction $A_{\rm V}$ is calculated using a prior on $E(B-V)$ and a reddening law parameterized by $R_{\rm V}$. Two different values of $R_{\rm V}$ are adopted in \cite{Hicken:2009dk} - $R_{\rm V} =3.1$ and $R_{\rm V} =1.7$. We note that there is some evidence that the MLCS treatment of colour introduces a systematic error into the determination of the absolute brightness and can bias cosmological parameter estimation. In particular, it has been argued that the `Hubble bubble' detected in \cite{Zehavi:1998gz,Conley:2007ng,Jha:2006fm} might be due to an incorrect assumption regarding the value of $R_{\rm V}$. For the purposes of this work we repeat our analysis using both the MLCS $R_{\rm V} = 1.7$ and SALT2 Constitution catalogs.

The recently constructed Lick Observatory Supernova Search (LOSS) sample was developed in \cite{Ganeshalingam:2013mia}. The data consists of $165$ $BVRI$ light curves of low redshift supernova, mainly sampled a week before maximum light in the B band. This set of objects was combined with data from the Calan/Tololo sample \cite{Hamuy:1996su}, CfA1-3 \cite{Riess:1998dv,Jha:2005jg,Hicken:2009df} and CSP \cite{Contreras:2009nt,Folatelli:2009nm}. For SN that are common to multiple surveys, the best sampled light curves are utilised. The SALT2 light curve fitting procedure was adopted to construct the distance modulus of $N_{\rm LOSS}=586$ data in the range $z=0.01-1.4$. Of the $226$ low-z supernova, $91$ are from LOSS and $45$ had distances published for the first time in \cite{Ganeshalingam:2013mia}. 

Finally, we use the combined catalog of ref.\cite{Johnson:2014kaa}. In \cite{Johnson:2014kaa} the aim was to construct a maximally homogeneous set of data points out to redshift $z < 0.07$. As such, the sample is drawn from a wide number of catalogs - specifically the data consists of $(40,128,135,58,33,26)$ SN from the LOSS, Tonry et al. \cite{Tonry:2003zg}, MLCS Constitution, Union, Kessler et al. \cite{Kessler:2009ys} and Carnegie Supernova Project \cite{Hamuy:2005tf,Folatelli:2009nm} samples respectively. When multiple measurements of the same object are known, the median value of the distance modulus is quoted. One arrives at a catalog containing $N=303$ SNIa, with distance errors of order $\sigma_{\rm d} \sim 5\%$. 

A number of small modifications to each catalog are made. Since we are testing for bulk flows of expected order $v_{\rm bulk} \sim 300-500 {\rm km s ^{-1}}$, we must have precise knowledge of the SN redshifts. We therefore update each data point using host galaxy information found in the NASA Extragalactic Database (NED)\footnote{http://ned.ipac.caltech.edu/}. We do not include any supernova in our analysis for which no host galaxy information is known. We also neglect any data for which the redshift error is larger than $\sigma_{\rm z} = 100 {\rm km s^{-1}}$. For the SN $2002{\rm hu}$, we do not use the redshift quoted in the NED database. Rather, we adopt a more recent measurement \cite{Blondin:2012ha}. Our use of the host galaxy redshifts is the primary reason for the small difference in results between this work and \cite{Colin:2010ds}.

The Constitution set adopt a different value of $H_{\rm 0,const}=65 {\rm km s^{-1} Mpc^{-1}}$ during the light curve fit procedure - for the purposes of combining data sets later we add the term $\delta \mu_{\rm h} = 5\log[H_{\rm 0, const}/H_{\rm 0,fid}]$ to each SN in this catalog, with $H_{\rm 0,fid} = 70$ ${\rm km s^{-1}}$ ${\rm Mpc^{-1}}$. All of the catalogs considered in this work introduce an error component in the distance modulus estimation to account for peculiar velocities, which is set at $\sigma_{\rm v} = 300{\rm km s^{-1}}$ in the Union 2.1 and LOSS set, and $\sigma_{\rm v} = 400{\rm km s^{-1}}$ for the Constitution sample. This contribution to the error is removed for the combined catalog of  \cite{Johnson:2014kaa}, and reintroduced in equation ($\ref{eq:ctot}$) via the $\sigma_{\rm v}^{2}$ contribution to the covariance. In ref.\cite{Johnson:2014kaa} $\sigma_{\rm v}$ was kept as a free parameter, and allowed to vary with the cosmological parameters. It was found that a value of $\sigma_{\rm v} = 400 {\rm km s^{-1}}$ was preferred by the data. We adopt this value whenever we use the combined catalog of \cite{Johnson:2014kaa}.

\section{\label{sec:4}Results}

We begin by calculating the $\Delta Q_{\rm d}$ values and their associated p-values for each supernova catalog, using realisation set A (that is, neglecting large scale velocity perturbations expected within the context of $\Lambda$CDM). For each catalog, we decompose the data into five unequally spaced redshift bins - see table \ref{tab:1} for their limits and number of SN contained within each bin. We perform our test using both concentric and cumulative redshift shells.

Our test function $Q(\theta,\phi,\delta)$ contains a free parameter $\delta$, which determines the width of the smoothing on the unit sphere. Decreasing values of $\delta$ will allow us to probe bulk velocity distributions in successively smaller regions of the sky, corresponding to local flows due to small scale over densities. However, the significance of such events is expected to be low due to the small number of participating SN Ia. Since our statistic $\Delta Q_{\rm d}$ was chosen to search for dipoles, and since a dipole signal would manifest itself as velocities coherent over entire hemispheres on the sky, we fix $\delta = \pi/2$ in what follows. 

We exhibit the $\Delta Q_{\rm d}$ $p$-value and direction of maximal anisotropy $(b_{\rm max},\ell_{\rm max})$ in table \ref{tab:1} for the Union 2.1, Constitution, LOSS and combined sets respectively. One can see consistent trends in all of the samples. The direction $(b_{\rm max},\ell_{\rm max})$ of maximal anisotropy in the cumulative redshift bins is consistent amongst the different data sets out to redshift $z < 0.06$. This is not a particularly surprising outcome, as the majority of low redshift SN are present in all four catalogs. However, it serves as a useful check that the previous detection of a bulk flow \cite{Colin:2010ds} is robust to both treatment of systematics and choice of light curve fitting procedure. However, the significance of any anisotropic signal is modest, with $p$-values typically greater than $p \sim 0.1$ in all catalogs and all redshift bins. Notable exceptions are the Union 2.1 and Combined samples, in cumulative redshift shell $0.015 < z < 0.045$, where the anisotropic signal is largest. 

We can state that for each individual catalog, there is no statistical significance of a bulk flow in any concentric redshift bin. The lack of significance is a reflection of the fact that there are insufficient number of supernovae in the individual shells, and coherent motion in a particular region of the sky cannot be strongly distinguished from random velocities when there are only a small number of data points. 

It is of interest to compare the results of the Constitution set, analysed using the two different light curve fitting procedures. One can see broad consistency in both the directions $(b_{\rm max},\ell_{\rm max})$ and significance when applying the SALT II and MLCS light curve fitting procedures to the same data sets. Both give comparable $p$-values and $(b_{\rm max},\ell_{\rm max})$ directions for the majority of the redshift bins considered. The only interesting deviation occurs in the redshift bin $0.045 \leq z < 0.06$, in which the MLCS procedure assigns considerably higher significance to the anisotropy than the SALT II fit. This bin is of interest, and we discuss it in more detail.

\begin{widetext}
\begin{center}
\begin{table}
\begin{tabular}{c|c|c|c|c|c}
\toprule 
 Catalog  & $0.015 \leq z < 0.025$ & $0.025 \leq z < 0.035$ & $0.035 \leq z < 0.045$ & $0.045 \leq z < 0.06$ & $0.06 \leq z < 0.1$  \\
\hline   \midrule
Union $2.1$       & 61 & 51 & 15 & 17 & 19  \\
Constitution      & 53 & 40 & 11 & 12 & 8  \\
LOSS              & 76 & 64 & 23 & 17 & 19  \\
Combined          & 98 & 67 & 22 & 27 & 12  \\
    \bottomrule
\end{tabular}
\caption{The number of supernova used in this work in each redshift bin, for each of the four catalogs adopted.}
\label{tab:1}
\end{table}
\end{center}
\end{widetext}

In the $0.045 \leq z < 0.06$ bin one can clearly see some consistency amongst the anisotropic directions $(b_{\rm max},\ell_{\rm max})$ in the Union and Constitution samples  - what appears to be a turnaround in direction relative to the $z < 0.045$ bins. This turnaround is most significant for the Constitution set, using the MLCS $R_{\rm V} = 1.7$ light curve fit. This behaviour was first observed in \cite{Colin:2010ds}, and was attributed to infall into the Shapley supercluster. However it was argued in \cite{Feindt:2013pma} that the observed turnaround was primarily due to the supernovae ${\rm SN}1995{\rm ac}$ and ${\rm SN}2003{\rm ic}$, that are located at almost maximal distance from Shapley on the sky. Hence the effect may not be due to this supercluster, but instead either a random alignment of two large residuals in a particular region of the sky, or infall into a different local overdensity. When constructing our residuals, we also find that the change in anisotropic direction in the $0.045 \leq z < 0.06$ bin is due to the ${\rm SN}1995{\rm ac}$ and ${\rm SN}2003{\rm ic}$ data points - when removing them we find that the direction of maximal isotropy changes to $(b_{\rm max},\ell_{\rm max})=(-14^{\circ},90^{\circ})$ with insignificant $p$-value $p = 0.492$ for the MLCS17 catalog. One can conclude that there is no evidence of back-infall into Shapley in the bin $0.045 \leq z < 0.06$. The fact that the MLCS light curve fit yields a higher significance than SALT II in this redshift bin is simply due to the fact that the MLCS approach estimates a considerably smaller distance modulus uncertainty for the offending SNe ${\rm SN}1995{\rm ac}$ and ${\rm SN}2003{\rm ic}$ - the observational uncertainty $\sigma_{\mu}$ is of order $\sim 30\%$ and $\sim 15\%$ smaller respectively when using MLCS compared to SALT II. Although current results show no evidence of infall into Shapley, more data is required in this redshift bin to definitively answer the question.

There is some evidence of a dipole in the cumulative redshift bins. The direction of the anisotropy remains qualitatively consistent in all cumulative bins and catalogs. The minimum $p$-value is observed in the $0.015 \leq z < 0.045$ bin and combined and Union catalogs. Beyond $z=0.045$, the significance of the detection drops due to two factors - the small number of supernova in the higher redshift bins and the effect of peculiar velocities becoming insignificant at high redshift (causing a decrease in the signal to noise).

\begin{widetext}
\begin{center}
\begin{table}
\begin{tabular}{c|l|ccc|c|l|ccc}
\toprule 
 $\Delta z$ & \multicolumn{1}{|c|}{Catalog}  & $b_{\rm max}$ & $\ell_{\rm max}$ & $p$  & $\Delta z$ & \multicolumn{1}{|c|}{Catalog} & $b_{\rm max}$ & $\ell_{\rm max}$ & $p$  \\
 \hline   \midrule
                       & Union 2.1  & $49^{\circ}$ & $259^{\circ}$ & $0.084$ &  & Union 2.1 & $49^{\circ}$ & $259^{\circ}$ & $0.084$    \\
                       & Const (SALT II)   & $20^{\circ}$ & $284^{\circ}$ & $0.624$   & & Const (SALT II)  & $20^{\circ}$ & $284^{\circ}$ & $0.624$    \\
$0.015 \leq z < 0.025$ & Const (MLCS 17)   & $67^{\circ}$ & $241^{\circ}$ & $0.692$  & $0.015 < z \leq 0.025$ & Const (MLCS 17) & $67^{\circ}$ & $241^{\circ}$ & $0.692$   \\
                       & LOSS    & $4^{\circ}$ & $247^{\circ}$ & $0.412$  &  & LOSS & $4^{\circ}$ & $247^{\circ}$ & $0.412$   \\
                       & Combined   & $27^{\circ}$ & $241^{\circ}$ & $0.179$  &  & Combined & $27^{\circ}$ & $241^{\circ}$ & $0.179$   \\
 & & & & & & & & & \\
                       & Union 2.1  & $-29^{\circ}$ & $289^{\circ}$ & $0.665$   &  & Union 2.1 & $29^{\circ}$ & $274^{\circ}$ & $0.166$    \\
                       & Const (SALT II)   & $36^{\circ}$ & $320^{\circ}$ & $0.271$  & & Const (SALT II)  & $27^{\circ}$ & $322^{\circ}$ & $0.201$   \\
$0.025 \leq z < 0.035$ & Const (MLCS 17)   & $40^{\circ}$ & $313^{\circ}$ & $0.202$  & $0.015 < z \leq 0.035$ & Const (MLCS 17) & $52^{\circ}$ & $288^{\circ}$ & $0.201$    \\
                       & LOSS    & $38^{\circ}$ & $320^{\circ}$ & $0.156$   &  & LOSS  & $39^{\circ}$ & $283^{\circ}$ & $0.177$   \\
                       & Combined   & $56^{\circ}$ & $328^{\circ}$ & $0.339$   &  & Combined & $41^{\circ}$ & $266^{\circ}$ & $0.119$    \\
 & & & & & & & & & \\
                       & Union 2.1  & $27^{\circ}$ & $320^{\circ}$ & $0.172$    &  & Union 2.1  & $31^{\circ}$ & $284^{\circ}$ & $0.063$     \\
                       & Const (SALT II)   & $25^{\circ}$ & $306^{\circ}$ & $0.672$   & & Const (SALT II)  & $27^{\circ}$ & $301^{\circ}$ & $0.123$    \\
$0.035 \leq z < 0.045$ & Const (MLCS 17)   & $36^{\circ}$ & $316^{\circ}$ & $0.192$  & $0.015 < z \leq 0.045$ & Const (MLCS 17) & $49^{\circ}$ & $299^{\circ}$ & $0.083$   \\
                       & LOSS    & $-27^{\circ}$ & $292^{\circ}$ & $0.534$   &  & LOSS & $20^{\circ}$ & $284^{\circ}$ & $0.149$     \\
                       & Combined   & $11^{\circ}$ & $313^{\circ}$ & $0.381$   &  & Combined & $38^{\circ}$ & $276^{\circ}$ & $0.070$    \\
 & & & & & & & & & \\
                       & Union 2.1   & $-49^{\circ}$ & $58^{\circ}$ & $0.412$   &  & Union 2.1  & $25^{\circ}$ & $295^{\circ}$ & $0.198$    \\
                       & Const (SALT II)   & $-54^{\circ}$ & $55^{\circ}$ & $0.572$  & & Const (SALT II)  & $22^{\circ}$ & $310^{\circ}$ & $0.216$    \\
$0.045 \leq z < 0.06$ & Const (MLCS 17)   & $-59^{\circ}$ & $68^{\circ}$ & $0.074$   & $0.015 < z \leq 0.06$ & Const (MLCS 17) & $38^{\circ}$ & $315^{\circ}$ & $0.372$    \\
                       & LOSS    & $54^{\circ}$ & $3^{\circ}$ & $0.457$   &  & LOSS & $22^{\circ}$ & $288^{\circ}$ & $0.159$     \\
                       & Combined   & $-12^{\circ}$ & $94^{\circ}$ & $0.495$   &  & Combined & $39^{\circ}$ & $281^{\circ}$ & $0.176$   \\
 & & & & & & & & & \\
                       & Union 2.1  & $-5^{\circ}$ & $43^{\circ}$ & $0.426$  &  & Union  2.1 & $25^{\circ}$ & $306^{\circ}$ & $0.295$  \\
                       & Const (SALT II)   & $54^{\circ}$ & $32^{\circ}$ & $0.574$   & & Const (SALT II)  & $27^{\circ}$ & $317^{\circ}$ & $0.197$   \\
$0.06 \leq z < 0.1$ & Const (MLCS 17)   & $-4^{\circ}$ & $65^{\circ}$ & $0.352$  & $0.015 < z \leq 0.1$ & Const (MLCS 17) & $41^{\circ}$ & $342^{\circ}$ & $0.431$    \\
                       & LOSS    & $52^{\circ}$ & $349^{\circ}$ & $0.532$   &  & LOSS & $27^{\circ}$ & $295^{\circ}$ & $0.114$   \\
                       & Combined   & $-54^{\circ}$ & $65^{\circ}$ & $0.788$  &  & Combined & $36^{\circ}$ & $280^{\circ}$ & $0.270$    \\
 & & & & & & & & & \\
    \bottomrule
\end{tabular}
\caption{The $p$-values and direction of maximal anisotropy detected with the residual method elucidated in section \ref{sec:2}, for five data sets and ten redshift bins - five concentric [left three columns] and five cumulative [right three columns]. Small $p$-values $p < 0.1$ indicate some deviation from the null isotropic hypothesis. One can observe little evidence of anisotropy in the concentric redshift shells, due to the modest number of supernova. The cumulative shells present a stronger case for anisotropy, with smallest $p$-values in the redshift shell $0.015 \leq z < 0.045$. We note that our use of the host galaxy redshifts is the primary reason for the small difference in results between this work and \cite{Colin:2010ds}.}
\label{tab:2}
\end{table}
\end{center}
\end{widetext}

\subsection{\label{sec:6}Consistency with $\Lambda$CDM}

Thus far we have been restricting ourselves to the set of realisations A, in which the large scale velocity perturbations expected within the $\Lambda$CDM model are neglected. We now turn our attention to the question of whether the anisotropic signal observed is significant within the context of standard cosmology. 

To do so, we take the combined catalog and repeat our analysis, constructing a $\Delta Q_{\rm d}$ probability distribution from realisation set B. We only perform our test for the combined catalog, as it contains the largest number of data points that are also relatively homogeneously distributed on the sky. We perform our test using the five cumulative shells, which are the only bins that exhibit any hints of an anisotropic signal when we compare the data to realisation set A. We note that here we have neglected cross correlations in the observational errors, i.e. systematic effects. Hence one should consider our results to be an upper bound on the significance of the anisotropic signal. Constructing a covariance matrix that accounts for systematics across different sub-catalogs, each using different light curve fitting procedures, is beyond the scope of this work. 

In table \ref{tab:3} we exhibit the results from the cumulative redshift bins, with the third column denoting the $p$-values obtained from realisation set B (the $p$-values obtained in the previous section are included for reference in column two). There is a very clear degradation in the p-value in all redshift bins when comparing the data to realisations in which large scale velocities expected within $\Lambda$CDM are consistently accounted for. One can conclude that when using realisation set B, there is absolutely no hint that the $\Lambda$CDM null hypothesis, on which the realisations are built, is in tension with the data. The local bulk flow is consistent with $\Lambda$CDM.

\begin{center}
\begin{table}
\begin{tabular}{c|cc}
\toprule 
 $\Delta z$ & $p_{\rm A}$ & $p_{\rm B}$ \\
 \hline   \midrule
 & & \\
$0.015 \leq z < 0.025$ & 0.179 & 0.371 \\ 
 & & \\
$0.015 \leq z < 0.035$ & 0.119 & 0.355 \\ 
 & & \\
$0.015 \leq z < 0.045$ & 0.070 & 0.290 \\ 
 & & \\
$0.015 \leq z < 0.060$ & 0.176 & 0.412 \\ 
 & & \\
$0.015 \leq z < 0.100$ & 0.270 & 0.531 \\ 
    \bottomrule
\end{tabular}
\caption{The $p$-values obtained when comparing the data value of $\Delta Q_{\rm d}$ to mock realisations set A [Column two] and B [Column three]. We are using the combined catalog only and adopt a value of $\delta = \pi/2$. One can clearly observe a significant degradation of statistical significance when we test the anisotropic signal against the $\Lambda$CDM model. Indeed, there is no significant evidence that the bulk flow is anomalous within the context of $\Lambda$CDM.}
\label{tab:3}
\end{table}
\end{center}

\subsection{\label{sec:5}Effects due to inhomogeneous data}

The method adopted in this work yields the $p$-value for the test function $\Delta Q_{\rm d}$, and the direction of maximal anisotropy which we denote $(b_{\rm max}, \ell_{\rm max})$. We now consider how the direction of maximal anisotropy inferred by the smoothed residual method might be biased due to an inhomogeneous distribution of data points. 

A bulk velocity would manifest itself as a directional dependent signal in the distance moduli, specifically a dipole in the luminosity distance. The magnitude of the effect on any given data point will depend on both its position relative to the bulk flow direction and its redshift, and hence the direction of maximal anisotropy detected by our method will be affected by both the anisotropic and inhomogeneous nature of the data distribution on the sky.

To test how well one can reproduce the direction of an underlying anisotropic signal, we create mock realisations of the combined catalog in which we keep the data positions and redshifts fixed and introduce an artificial bulk flow. Specifically, the luminosity distance of the $i^{\rm th}$ data point is estimated as \cite{Bonvin:2006en}

\begin{equation}\label{eq:bv} d_{\rm L,i} = {(1+z_{\rm i})c \over H_{0}} \int_{0}^{z_{\rm i}} {d\bar{z} \over h(\bar{z})} + {V_{\rm bulk} (1+z_{\rm i})^{2} \over H(z_{\rm i})} \cos[\theta_{\rm i}] \end{equation}

\noindent where $V_{\rm bulk}$ is the magnitude of the bulk velocity, and $\theta_{\rm i}$ is the angle sub-tending the direction of the $i^{\rm th}$ supernova on the sky and the direction of the bulk velocity. The distance modulus of each data point is then constructed using this luminosity distance, and Gaussian noise is added to each point, using the square of the catalog observational uncertainty as the variance. We generate $N_{\rm real} = 15000$ realisations of the data - for each realisation we input a bulk velocity into the luminosity distance ($\ref{eq:bv}$) of magnitude $V_{\rm bulk} = 400 {\rm km s^{-1}}$ and with a random direction on the sphere - $(b_{\rm v},\ell_{\rm v})$. We then calculate the difference between the actual bulk velocity direction $(b_{\rm v},\ell_{\rm v})$ and the reconstructed direction $(b_{\rm max},\ell_{\rm max})$ that yields a maximum value of $\Delta Q_{\rm D}$, for each realisation. In the following analysis, we only keep realisations in which there is a `significant' detection of the inputted dipole, that is those with a $p$-value $p < 0.05$ relative to simulations in which $V_{\rm bulk} = 0$. 

\begin{figure*}
\centering
\mbox{\resizebox{0.45\textwidth}{!}{\includegraphics[angle=0]{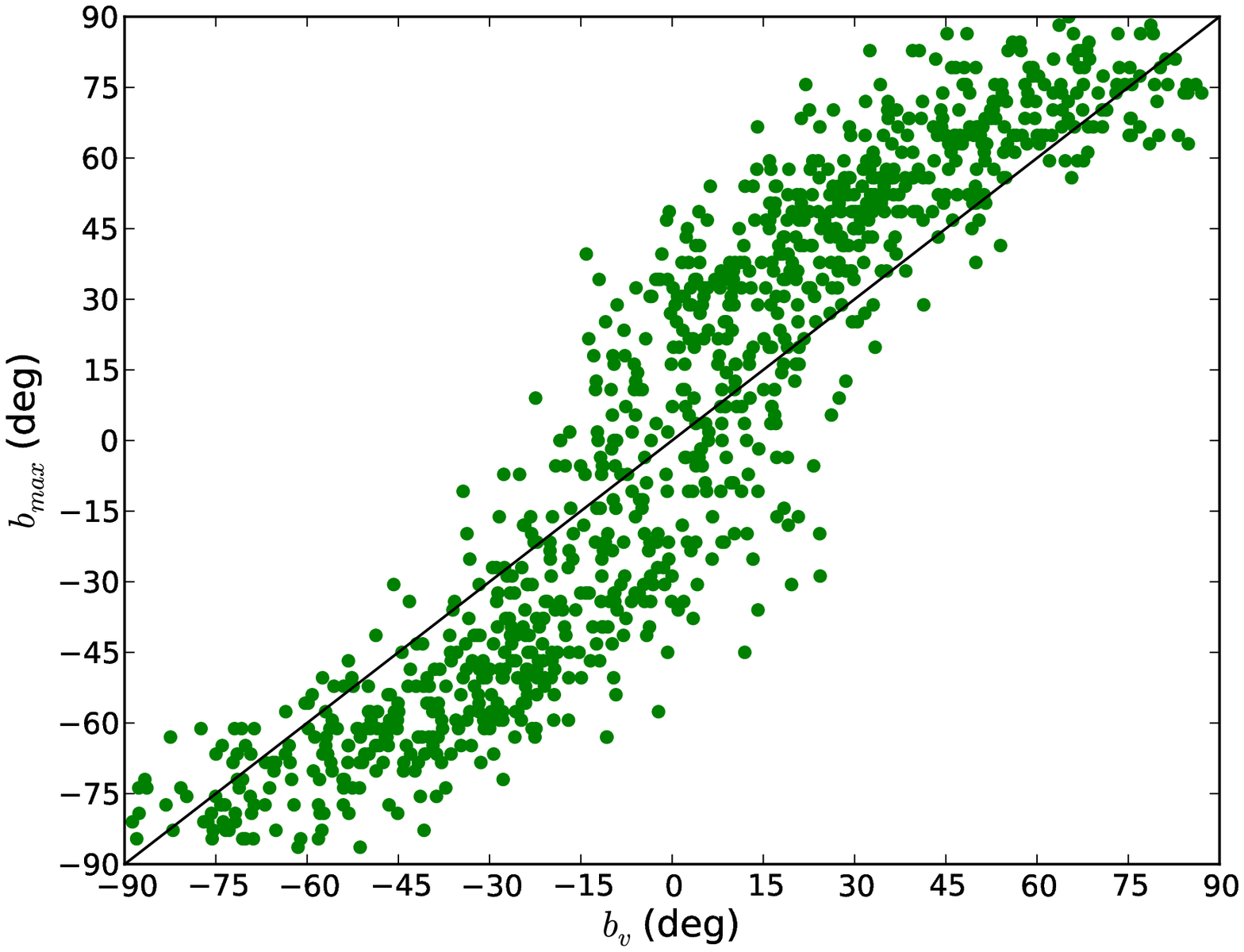}}}
\mbox{\resizebox{0.45\textwidth}{!}{\includegraphics[angle=0]{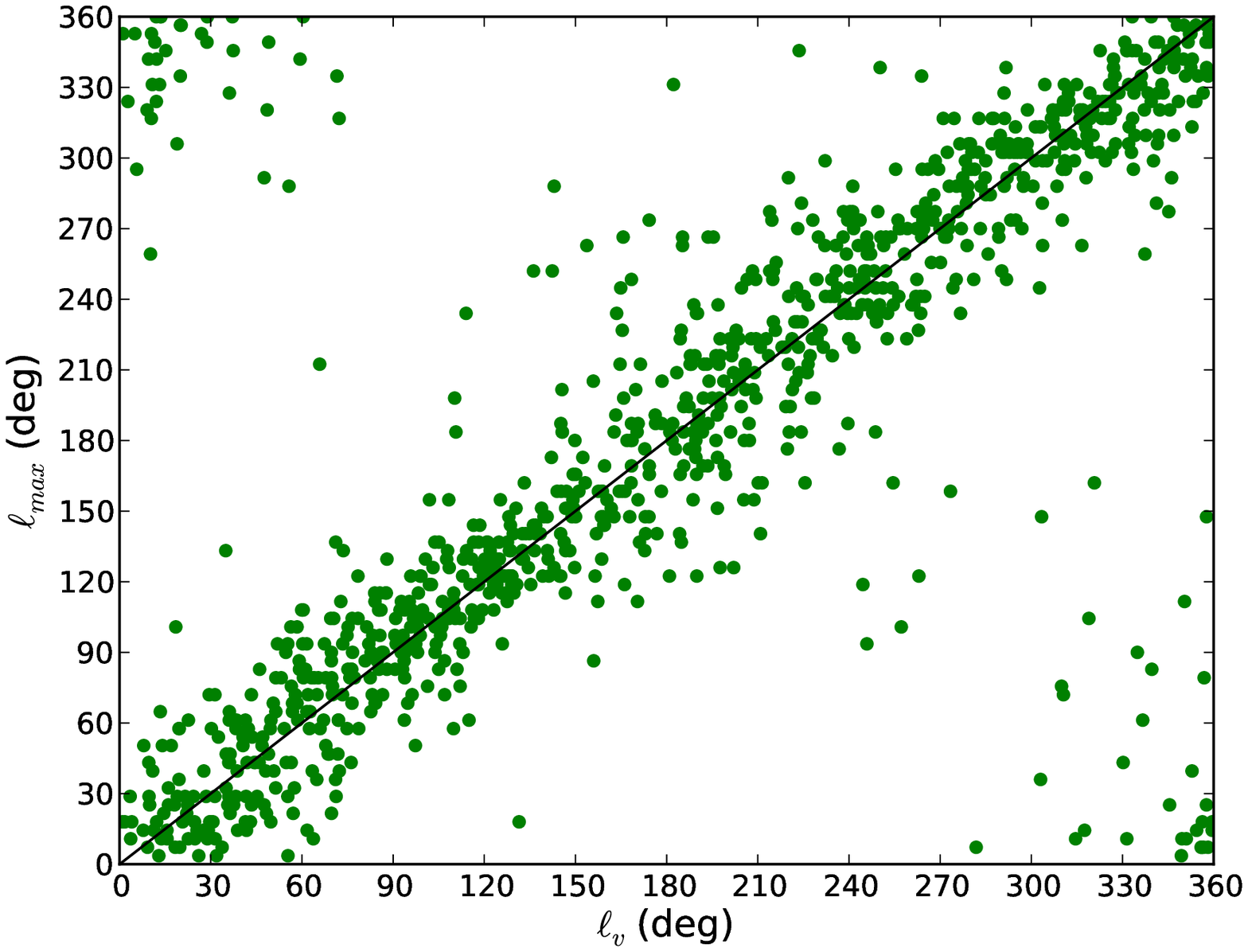}}}
\caption{We exhibit the scatter between the bulk velocity direction $b_{\rm v}$ and $\ell_{\rm v}$ inputted into our mock data set and the reconstructed direction $b_{\rm max}$, $\ell_{\rm max}$ obtained using our smoothing method. We only show an indicative subsample of the realisations that yield a significant detection of the dipole, that is those realisations with $p$-value $p < 0.05$ relative to $V_{\rm bulk}=0$ simulations. The galactic latitude and longitude are exhibited in the left and right panels respectively. The scatter in the longitude is symmetric around $\ell_{\rm max} = \ell_{\rm v}$, exhibited as a solid black line, and is therefore consistent with no bias. However the reconstructed latitude $b_{\rm max}$ is consistently higher(lower) than $b_{\rm v}$ in the north(south) galactic plane, indicating a bias away from the galactic center. This result is due to the sparsity of data in the region $|b| < 20^{\circ}$, and will be a generic feature of any attempt to estimate the direction of an anisotropic signal given inhomogeneous data. We note that the scatter in the top left and bottom right corners of the $(\ell_{\rm max},\ell_{\rm v})$ plot are due to the periodicity of the longitude $\ell \to \ell + 360^{\circ}$. }
\label{fig:2} 
\end{figure*}

We perform our test using the combined catalog and redshift bin $0.015 \leq z < 0.045$, which constitutes the redshift bin in which an anisotropic signal has the highest significance. For each realisation we randomize the direction of the dipole isotropically on the unit sphere, with no priors imposed on $(b_{\rm v},\ell_{\rm v})$. In fig.\ref{fig:2} we exhibit the scatter between $(b_{\rm v},b_{\rm max})$ and $(\ell_{\rm v},\ell_{\rm max})$ for the realisations. For a homogeneous data set, where there would be no bias, the scatter in fig.\ref{fig:2} would be centered on $b_{\rm v} = b_{\rm max}$ and $\ell_{\rm v} = \ell_{\rm max}$, denoted as black lines in the figures. Indeed, one can observe exactly such behaviour in the $(\ell_{\rm max},\ell_{\rm v})$ plot. However, there is a clear bias in the galactic latitude, with the method reconstructing a $b_{\rm max}$ that is consistently larger(smaller) than $b_{\rm v}$ in the north(south) galactic hemisphere. This is a reflection of the fact that the data is sparse in the region $|b| < 20^{\circ}$, and the method will preferentially select a region of the sky where the data is prevalent. 

We note two peculiarities in the $(\ell_{\rm v},\ell_{\rm max})$ scatter. The concentration of points in the top left and bottom right of the plot are due to the periodic nature of the longitude coordinate $\ell \to \ell + 360^{\circ}$, with low $\ell$ points scattering to $\ell \sim 360^{\circ}$ and vice versa. The small number of points that are otherwise scattered far from the expected $\ell_{\rm max} = \ell_{\rm v}$ relation are predominantly those in which the underlying bulk velocity in the realisation was placed at $|b_{\rm v}| < 20^{\circ}$ - for these values the uncertainty on $\ell_{\rm v}$ increases dramatically due to the lack of data in this region.    

\begin{figure*}
\centering
\mbox{\resizebox{0.45\textwidth}{!}{\includegraphics[angle=0]{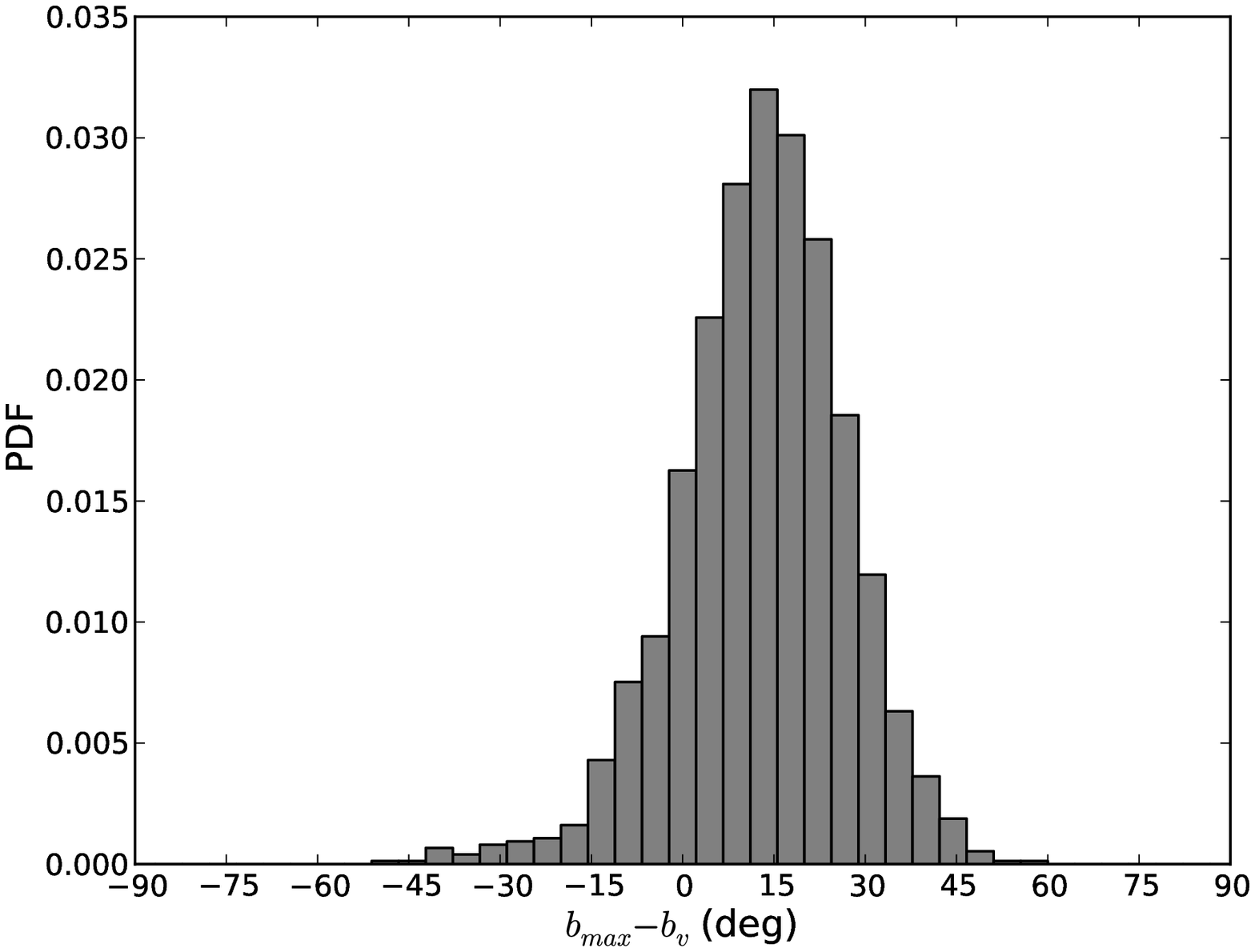}}}
\mbox{\resizebox{0.45\textwidth}{!}{\includegraphics[angle=0]{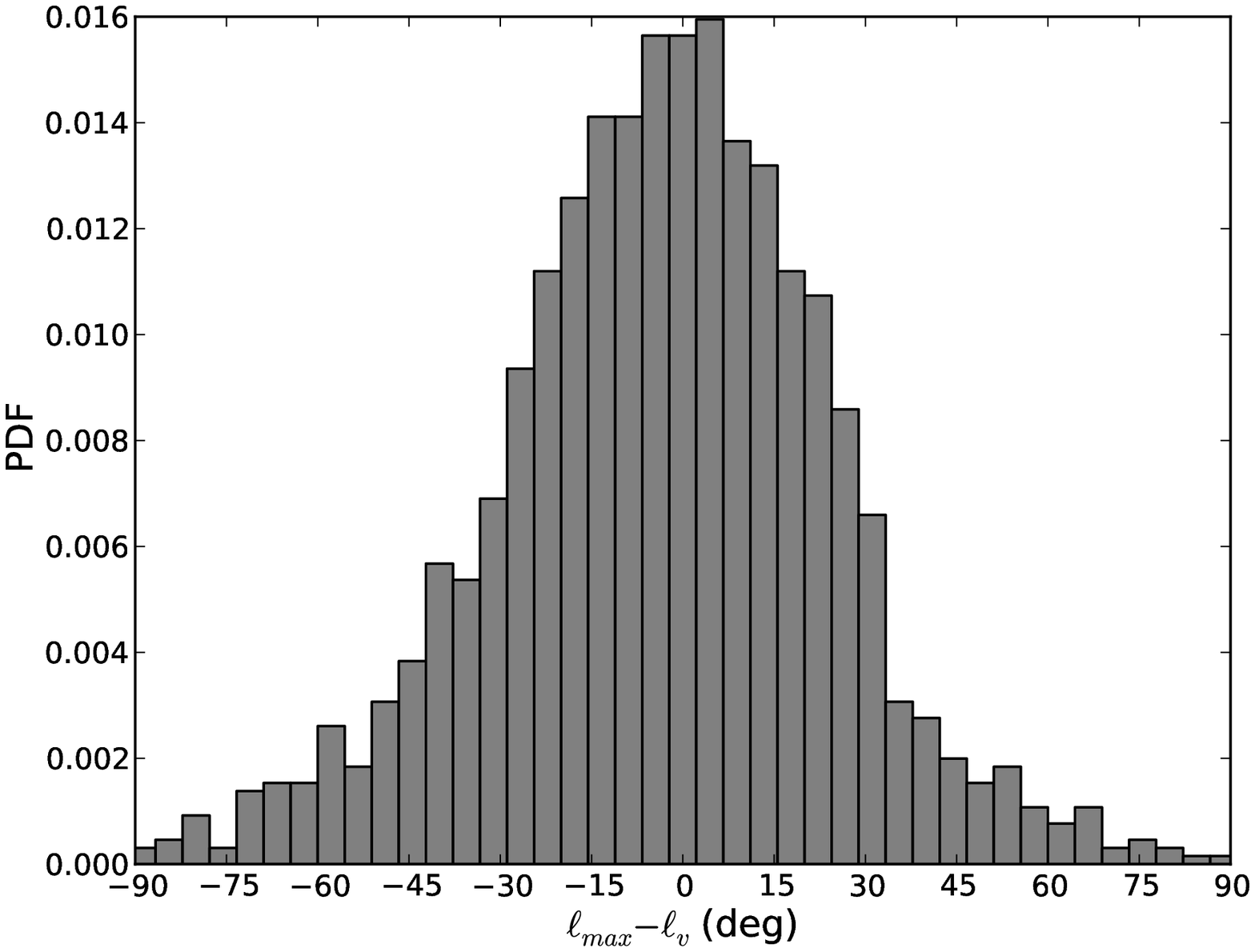}}}
\mbox{\resizebox{0.45\textwidth}{!}{\includegraphics[angle=0]{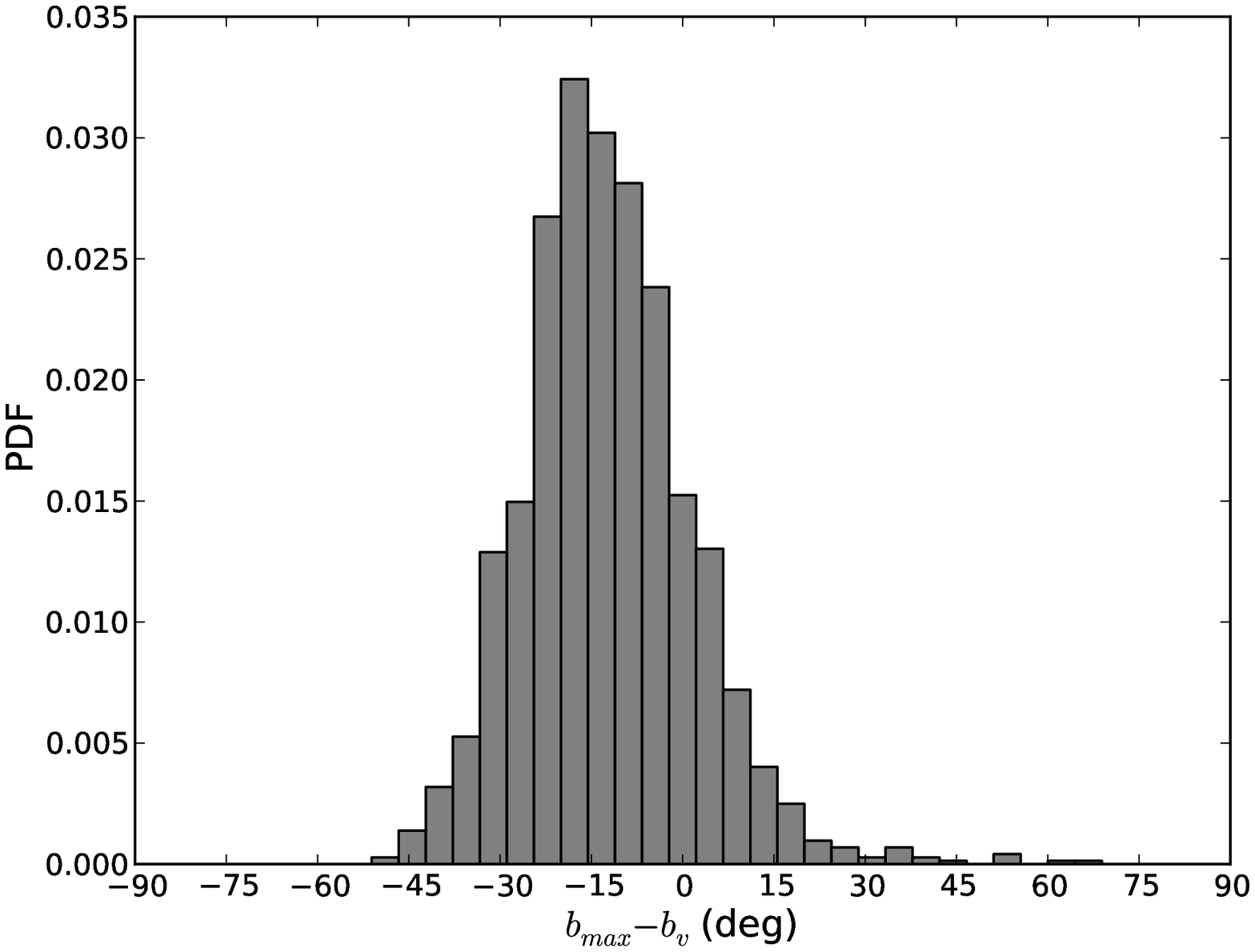}}}
\mbox{\resizebox{0.45\textwidth}{!}{\includegraphics[angle=0]{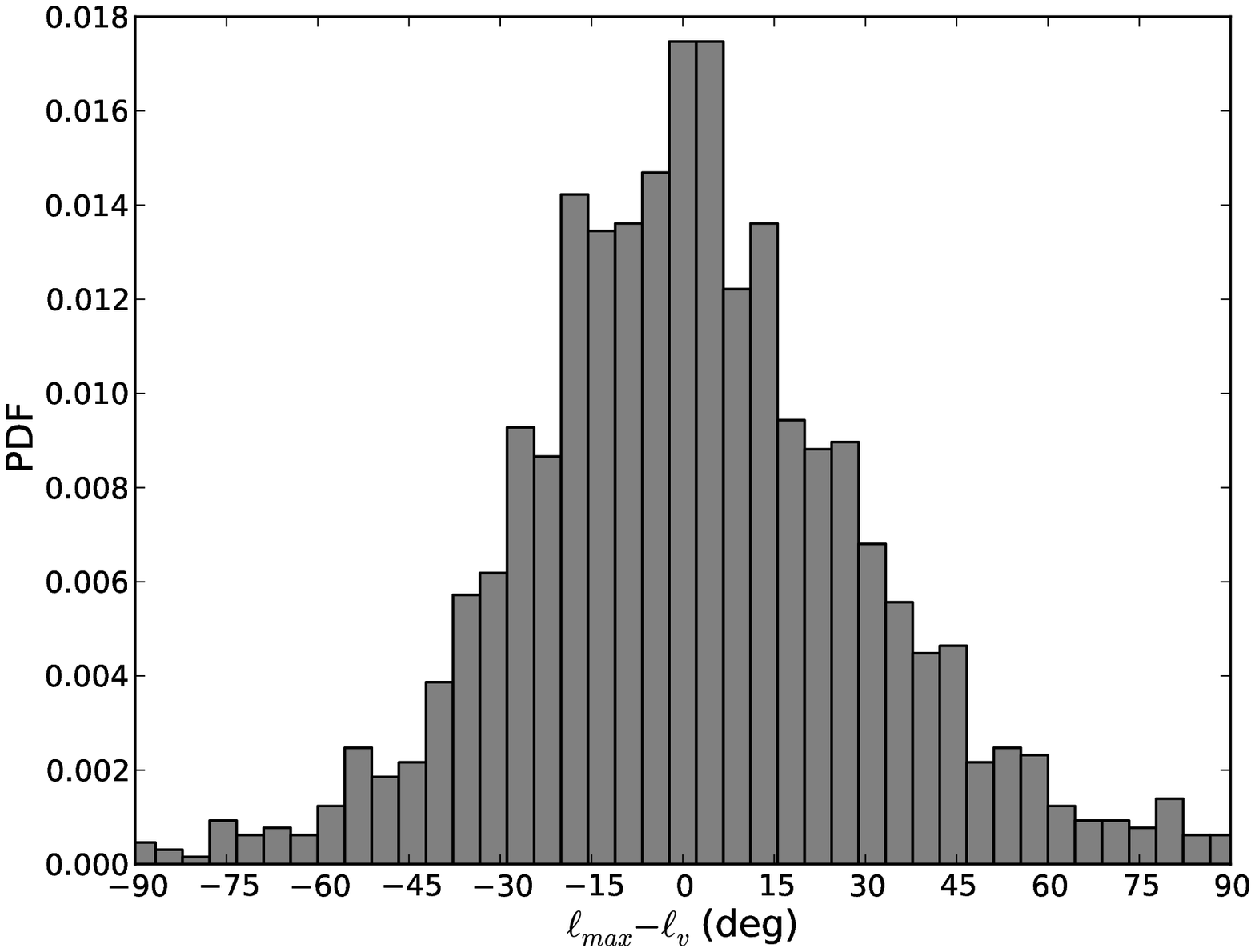}}}
\caption{We exhibit empirical probability distributions of $b_{\rm max} - b_{\rm v}$ (left panels) and $\ell_{\rm max} - \ell_{\rm v}$ (right panels) in the north (top panels) and south (bottom panels) galactic plane. We keep only realisations with $|b_{\rm v}| > 20^{\circ}$. The distributions are consistent in both the north and south galactic planes, with $\ell_{\rm max} - \ell_{\rm v}$ exhibiting no strong bias. The latitude is biased away from $b_{\rm max} = b_{\rm v}$ in both planes however.}
\label{fig:3} 
\end{figure*}

In fig.\ref{fig:3} we exhibit $(b_{\rm max}-b_{\rm v})$ and $(\ell_{\rm max}-\ell_{\rm v})$ as empirical probability distributions, where we keep only realisations for which $b_{\rm v} > 20^{\circ}$ (top panels) and $b_{\rm v} < -20^{\circ}$ (bottom panels). One can see that the galactic longitude yields distributions consistent with $\ell_{\rm v} = \ell_{\rm max}$, but the latitude reconstruction is biased. We estimate the bias $(\Delta b_{\rm max},\Delta \ell_{\rm max})$ in the north and south galactic planes by calculating the mean of the distributions in fig.\ref{fig:3}, and the uncertainty $(\delta b_{\rm max},\delta \ell_{\rm max})$ as the sample variance. We exhibit the values of $(\Delta b, \Delta \ell)$ and $(\delta b, \delta \ell)$ in the first row of table \ref{tab:4}.

We repeat our calculation using realisations in which a bulk velocity of magnitude $V_{\rm bulk} = 800 {\rm km s^{-1}}$ is introduced. In doing so, we obtain similar distributions to fig.\ref{fig:3}. We find that the mean values of the distributions do not vary appreciably when we increase $V_{\rm bulk}$, indicating that the bias $(\Delta b_{\rm max}, \Delta \ell_{\rm max})$ is not a strong function of the underlying bulk flow but rather an intrinsic feature of the data. However as one might expect the uncertainty (sample variance, $(\delta b_{\rm max}, \delta \ell_{\rm max})$)  of the distributions decreases, reflecting the signal to noise increase associated with a larger bulk flow. In table \ref{tab:4} we exhibit the values of the mean (bias) and sample variance (uncertainty) of the distributions in the north and south galactic planes, for both $V_{\rm bulk} = (400, 800) {\rm km s^{-1}}$ mock data sets.

\begin{center}
\begin{table}
\begin{tabular}{c|c|c|c|c|}
\toprule 
  & \multicolumn{2}{|c|}{North ($b_{\rm v} > 20^{\circ}$)}  & \multicolumn{2}{|c|}{South ($b_{\rm v} < -20^{\circ}$)}  \\
 \hline   \midrule
 $V_{\rm bulk} ({\rm km s^{-1}}) $   & $(\Delta b,\Delta \ell)$  & $(\delta b,\delta \ell)$ & $(\Delta b,\Delta \ell)$ & $(\delta b,\delta \ell)$   \\
 & & & &  \\
 400 & $(13^{\circ},-3^{\circ})$ & $(14^{\circ},28^{\circ})$ & $(-12^{\circ},2^{\circ})$ & $(14^{\circ},29^{\circ})$ \\ 
 & & & &  \\
 800 & $(15^{\circ},-4^{\circ})$ & $(9^{\circ},22^{\circ})$ & $(-13^{\circ},2^{\circ})$ & $(9^{\circ},21^{\circ})$ \\ 
    \bottomrule
\end{tabular}
\caption{We exhibit the bias $(\Delta b,\Delta \ell)$ and uncertainty $(\delta b,\delta \ell)$ in the reconstructed direction $(b_{\rm max},\ell_{\rm max})$ for realisations with $b_{\rm v} > 20^{\circ}$ (columns $(2,3)$) and $b_{\rm v} < -20^{\circ}$ (columns $(4,5)$). We adopt two sets of realisations with an inputted bulk velocity magnitude of $V_{\rm bulk} = 400 {\rm km s^{-1}}$ and $V_{\rm bulk} = 800 {\rm km s^{-1}}$. We note that the bias in galactic longitude $\ell$ is negligible, however the latitude $b_{\rm max}$ is systematically shifted away from the galactic plane in both the northern and southern hemispheres. The bias does not appreciably vary with $V_{\rm bulk}$, indicating that $\Delta b$ is due to the inhomogeneous nature of the data. The uncertainty $(\delta b, \delta \ell)$ decreases with increasing $V_{\rm bulk}$, as a result of increased signal to noise.}
\label{tab:4}
\end{table}
\end{center}

\begin{figure}
\centering
\mbox{\resizebox{0.45\textwidth}{!}{\includegraphics[angle=0]{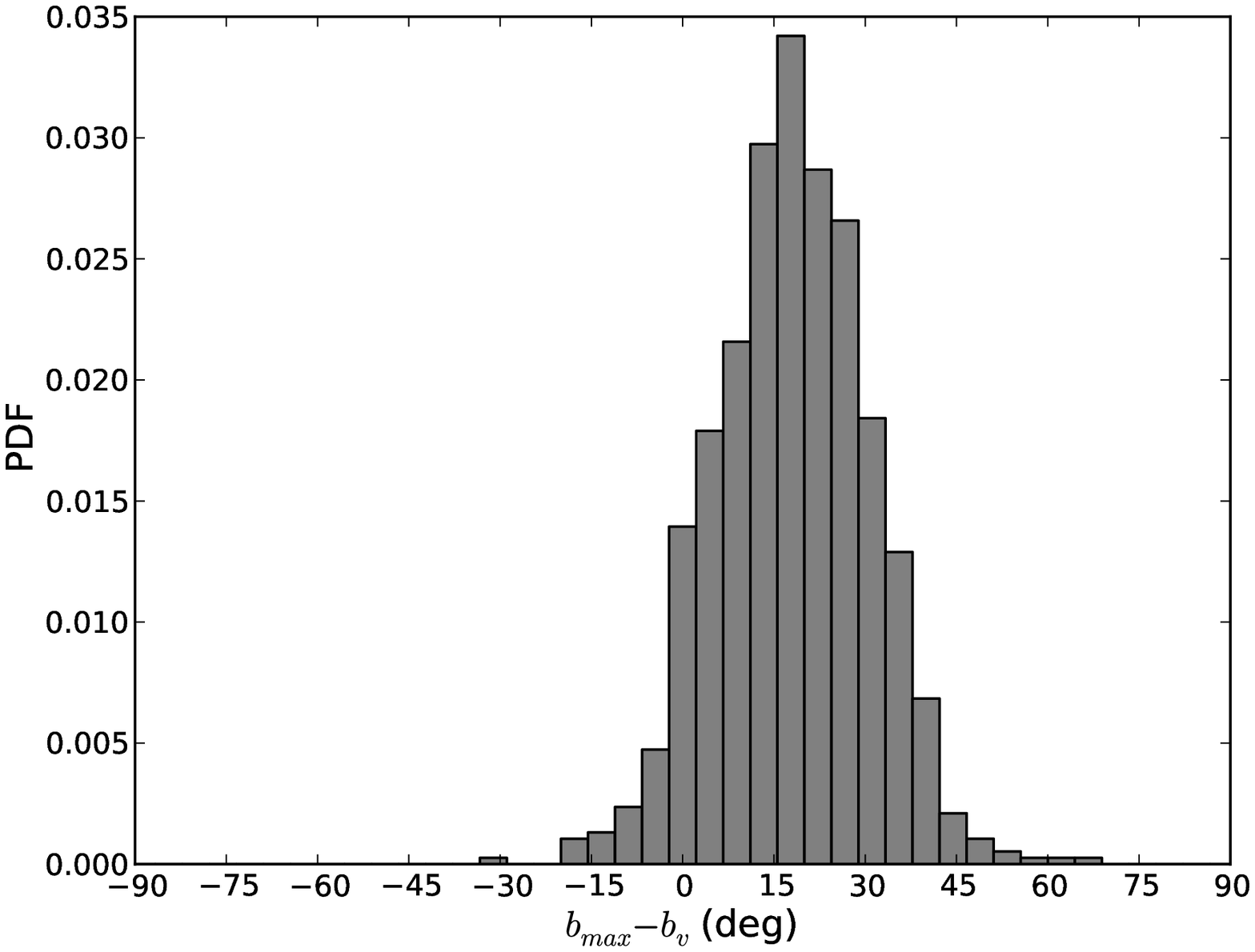}}}
\caption{We exhibit the distribution of $b_{\rm max} - b_{\rm v}$ for all realisations that have a reconstructed direction in the range $34^{\circ} < b_{\rm max} < 42^{\circ}$, regardless of the underlying direction $b_{\rm v}$. We use these realisations to estimate the bias and uncertainty in the $b_{\rm max}$ value obtained using the actual supernova data (combined catalog, redshift bin $0.015 < z < 0.045$).}
\label{fig:4} 
\end{figure}

Finally, we use our realisations to estimate the bias and uncertainty associated with the measurement of $(b_{\rm max},\ell_{\rm max})$ obtained using the data - specifically the $0.015 < z < 0.045$ redshift shell of the combined catalog. To do so we select the subset of $V_{\rm bulk} = 400 {\rm km s^{-1}}$ realisations that have a {\it reconstructed} direction $b_{\rm max}$ in the range $34^{\circ} < b_{\rm max} < 42^{\circ}$, regardless of the actual bulk velocity latitude $b_{\rm v}$ (recall that the data yielded a direction of maximal anisotropy $b_{\rm max} = 38^{\circ}$ - see table \ref{tab:2}). We construct a distribution of $(b_{\rm max} - b_{\rm v})$ of this subset, which we exhibit in fig.\ref{fig:4}. We estimate the bias $\Delta b_{\rm max}$ and uncertainty $\delta b_{\rm max}$ on our data measurement $b_{\rm max} = 38^{\circ}$ as the mean and sample variance of this distribution respectively. We find $\Delta b_{\rm max} = 18^{\circ}$ and $\delta b_{\rm max} = 12^{\circ}$. As the longitude $\ell_{\rm max}$ has been shown to be unbiased, we take $\Delta \ell_{\rm max} = 0^{\circ}$ and $\delta \ell_{\rm max} = 29^{\circ}$ from the whole sample. Hence for the combined catalog and redshift bin  $0.015 < z < 0.045$, we estimate the direction of maximal anisotropy to be $(b_{\rm max},\ell_{\rm max}) = (20^{\circ},276^{\circ}) \pm (12^{\circ},29^{\circ})$. We note that the final reconstructed direction is in agreement with various theoretical and observational works \cite{Campanelli:2009tk,Gibelyou:2012ri,Feindt:2013pma}. Our result is also in qualitative agreement with recent work by the authors \cite{Appleby:2014lra}, where a non-parametric reconstruction of the bulk flow direction was undertaken using the local galaxy distribution and the associated luminosity function.

Although our analysis raises concern as to the reliability of the $b_{\rm max}$ values presented in table \ref{tab:2}, one should recall that the method is primarily designed as a null hypothesis test. The principle result of the method are the $p$-values obtained for each catalog and redshift bin - small values indicate any kind of violation of the isotropic hypothesis. One can make no strong claim as to the direction or underlying cause of anisotropy when performing a null test for the current data sets - we can simply state whether the data is consistent with the null hypothesis. However by using mock data sets and making assumptions as to the nature of the anisotropy, we are able to make a quantitative statement regarding the accuracy of our recovered $(b_{\rm max},\ell_{\rm max})$ directions. Clearly, both the bias and the uncertainty in $(b_{\rm max},\ell_{\rm max})$ will depend upon the data set used.

\section{\label{sec:7}Discussion}

In this work we have used a non-parametric method to test the underlying assumption of isotropy in the low redshift supernova data sets. Our first test, using realisation set A, is tantamount to the question - given the statistical, systematic and astrophysical uncertainties associated with the supernova measurements, are the data consistent with flat, isotropic $\Lambda$CDM expansion with zero bulk flow? Our answer to this question is that there is a hint in the Union and combined catalogs of an anisotropic signal, which is most pronounced in the $0.015 < z < 0.045$ redshift bin, with $p$-value $p=0.07$. Due to the small number of data points, tomography yields no significant findings in any redshift bin. When we compare the data to realisation set B, where large scale velocity perturbations within the context of $\Lambda$CDM are accounted for, we find that there is no hint of any anomalous deviation from the null hypothesis, with the smallest $p$-value $p=0.29$. One can conclude that there is some evidence of a dipole in the supernova catalogs, however this bulk flow is consistent with the standard cosmology.

We attempted to use the method of smoothed residuals to estimate the direction of the bulk flow, however the reconstructed direction will be biased by the inhomogeneous distribution of data on the sky. To estimate this effect, we created mock realisations of the data in which we inserted a bulk flow of magnitude $V_{\rm bulk} = 400 {\rm km s^{-1}}$ and random direction $(b_{\rm v}, \ell_{\rm v})$. We then used our method to attempt to reconstruct this direction. We found that the reconstructed galactic longitude $\ell_{\rm max}$ is consistent with the inputted value $\ell_{\rm v}$, however the recovered galactic latitude $b_{\rm max}$ is systematically biased away from the galactic plane. Such behaviour is expected given the sparsity of data in the region $|b| < 20^{\circ}$, and we expect that such a bias is unavoidable in reconstructions when the data is inhomogeneously distributed. After correcting for a bias of $(\Delta b, \Delta \ell) = (18^{\circ},0^{\circ})$, we quote the direction of maximal anisotropy as $(b_{\max},\ell_{\max}) = (20^{\circ}, 276^{\circ}) \pm (12^{\circ},29^{\circ})$ for the combined catalog and redshift shell $0.015 < z < 0.045$. 

We note that our result is qualitatively consistent with recent work \cite{Feindt:2013pma}. In \cite{Feindt:2013pma}, two approaches are used to conclude that the Union 2 sample contains a dipole contribution. The first of these approaches is an explicit dipole fit to the data - using equation ($\ref{eq:bv}$) and fitting for the magnitude of the bulk velocity $V_{\rm bulk}$ and its direction on the sky $(b_{\rm v},\ell_{\rm v})$. Our analysis does not preclude the existence of a detectable dipole - indeed both our approach and the dipole fit yield a qualitatively consistent direction $(b_{\rm max}, \ell_{\rm max})$ of maximal anisotropy once we have eliminated the bias in our estimation.

It is clear that there is some evidence of an anisotropic signal in the supernova data. However, more data and better control of the observational uncertainties are required before one can pin down the magnitude, direction and redshift extent of the local bulk flow. However, one can state that the local bulk flow observed in current SNe data is consistent with the $\Lambda$CDM model.

\acknowledgements{The authors would like to thank Subir Sakar, Chris Blake and Jeppe Nielsen for helpful discussions. S.A.A and A.S wish to acknowledge support from the Korea Ministry of Education, Science and Technology, Gyeongsangbuk-Do and Pohang City for Independent Junior Research Groups at the Asia Pacific Center for Theoretical Physics. A.S would like to acknowledge the support of the National Research Foundation of Korea (NRF-2013R1A1A2013795). A. J is supported by the Australian Research Council Centre of Excellence for All-Sky Astrophysics (CAASTRO) through project number CE110001020. }

\newpage

\bibliography{mybib}{}

\begin{thebibliography}{10}

\bibitem{Riess:1998cb}
A.~G. Riess {\em et~al.}, ``{Observational evidence from supernovae for an
  accelerating universe and a cosmological constant},'' {\em Astron.J.},
  vol.~116, pp.~1009--1038, 1998.

\bibitem{Perlmutter:1998np}
S.~Perlmutter {\em et~al.}, ``{Measurements of Omega and Lambda from 42 high
  redshift supernovae},'' {\em Astrophys.J.}, vol.~517, pp.~565--586, 1999.

\bibitem{Garnavich:1998th}
P.~M. Garnavich {\em et~al.}, ``{Supernova limits on the cosmic equation of
  state},'' {\em Astrophys.J.}, vol.~509, pp.~74--79, 1998.

\bibitem{Riess:1998dv}
A.~G. Riess, R.~P. Kirshner, B.~P. Schmidt, S.~Jha, P.~Challis, {\em et~al.},
  ``{BV RI light curves for 22 type Ia supernovae},'' {\em Astron.J.},
  vol.~117, pp.~707--724, 1999.

\bibitem{Blakeslee:2003dz}
J.~P. Blakeslee {\em et~al.}, ``{Discovery of two distant type Ia supernovae in
  the Hubble Deep Field North with the advanced camera for surveys},'' {\em
  Astrophys.J.}, vol.~589, pp.~693--703, 2003.

\bibitem{Tonry:2003zg}
J.~L. Tonry {\em et~al.}, ``{Cosmological results from high-z supernovae},''
  {\em Astrophys.J.}, vol.~594, pp.~1--24, 2003.

\bibitem{Riess:2003gz}
A.~G. Riess {\em et~al.}, ``{Identification of type Ia supernovae at redshift
  1.3 and beyond with the Advanced Camera for Surveys on HST},'' {\em
  Astrophys.J.}, vol.~600, pp.~L163--L166, 2004.

\bibitem{Barris:2003dq}
B.~J. Barris, J.~L. Tonry, S.~Blondin, P.~Challis, R.~Chornock, {\em et~al.},
  ``{23 High redshift supernovae from the IFA Deep Survey: Doubling the SN
  sample at $z > 0.7$},'' {\em Astrophys.J.}, vol.~602, pp.~571--594, 2004.

\bibitem{Matheson:2004dj}
T.~Matheson, S.~Blondin, R.~J. Foley, R.~Chornock, A.~V. Filippenko, {\em
  et~al.}, ``{Spectroscopy of high-redshift supernovae from the ESSENCE
  project: The First two years},'' {\em Astron.J.}, vol.~129, pp.~2352--2375,
  2005.

\bibitem{Hicken:2009dk}
M.~Hicken, W.~M. Wood-Vasey, S.~Blondin, P.~Challis, S.~Jha, {\em et~al.},
  ``{Improved Dark Energy Constraints from ~100 New CfA Supernova Type Ia Light
  Curves},'' {\em Astrophys.J.}, vol.~700, pp.~1097--1140, 2009.

\bibitem{Hicken:2009df}
M.~Hicken, P.~Challis, S.~Jha, R.~P. Kirsher, T.~Matheson, {\em et~al.},
  ``{CfA3: 185 Type Ia Supernova Light Curves from the CfA},'' {\em
  Astrophys.J.}, vol.~700, pp.~331--357, 2009.

\bibitem{Hamuy:2005tf}
M.~Hamuy, G.~Folatelli, N.~I. Morrell, M.~M. Phillips, N.~B. Suntzeff, {\em
  et~al.}, ``{The carnegie supernova project: the low-redshift survey},'' {\em
  Publ.Astron.Soc.Pac.}, vol.~118, pp.~2--20, 2006.

\bibitem{Folatelli:2009nm}
G.~Folatelli, M.~Phillips, C.~R. Burns, C.~Contreras, M.~Hamuy, {\em et~al.},
  ``{The Carnegie Supernova Project: Analysis of the First Sample of
  Low-Redshift Type-Ia Supernovae},'' {\em Astron.J.}, vol.~139, pp.~120--144,
  2010.

\bibitem{Astier:2005qq}
P.~Astier {\em et~al.}, ``{The Supernova legacy survey: Measurement of
  omega(m), omega(lambda) and W from the first year data set},'' {\em
  Astron.Astrophys.}, vol.~447, pp.~31--48, 2006.

\bibitem{Sako:2007ms}
M.~Sako {\em et~al.}, ``{The Sloan Digital Sky Survey-II Supernova Survey:
  Search Algorithm and Follow-up Observations},'' {\em Astron.J.}, vol.~135,
  pp.~348--373, 2008.

\bibitem{Miknaitis:2007jd}
G.~Miknaitis, G.~Pignata, A.~Rest, W.~Wood-Vasey, S.~Blondin, {\em et~al.},
  ``{The ESSENCE Supernova Survey: Survey Optimization, Observations, and
  Supernova Photometry},'' {\em Astrophys.J.}, vol.~666, pp.~674--693, 2007.

\bibitem{WoodVasey:2007jb}
W.~M. Wood-Vasey {\em et~al.}, ``{Observational Constraints on the Nature of
  the Dark Energy: First Cosmological Results from the ESSENCE Supernova
  Survey},'' {\em Astrophys.J.}, vol.~666, pp.~694--715, 2007.

\bibitem{Holtzman:2008zz}
J.~A. Holtzman {\em et~al.}, ``{The Sloan Digital Sky Survey-II Photometry and
  Supernova IA Light Curves from the 2005 Data},'' {\em Astron.J.}, vol.~136,
  pp.~2306--2320, 2008.

\bibitem{Kessler:2009ys}
R.~Kessler, A.~Becker, D.~Cinabro, J.~Vanderplas, J.~A. Frieman, {\em et~al.},
  ``{First-year Sloan Digital Sky Survey-II (SDSS-II) Supernova Results: Hubble
  Diagram and Cosmological Parameters},'' {\em Astrophys.J.Suppl.}, vol.~185,
  pp.~32--84, 2009.

\bibitem{Freedman:2009vv}
W.~L. Freedman, C.~R. Burns, M.~Phillips, P.~Wyatt, S.~Persson, {\em et~al.},
  ``{The Carnegie Supernova Project: First Near-Infrared Hubble Diagram to
  z~0.7},'' {\em Astrophys.J.}, vol.~704, pp.~1036--1058, 2009.

\bibitem{Ganeshalingam:2013mia}
M.~Ganeshalingam, W.~Li, and A.~V. Filippenko, ``{Constraints on dark energy
  with the LOSS SN Ia sample},'' 2013.

\bibitem{Amanullah:2010vv}
R.~Amanullah, C.~Lidman, D.~Rubin, G.~Aldering, P.~Astier, {\em et~al.},
  ``{Spectra and Light Curves of Six Type Ia Supernovae at $0.511 < z < 1.12$
  and the Union2 Compilation},'' {\em Astrophys.J.}, vol.~716, pp.~712--738,
  2010.

\bibitem{Suzuki:2011hu}
N.~Suzuki, D.~Rubin, C.~Lidman, G.~Aldering, R.~Amanullah, {\em et~al.}, ``{The
  Hubble Space Telescope Cluster Supernova Survey: V. Improving the Dark Energy
  Constraints Above $z > 1$ and Building an Early-Type-Hosted Supernova
  Sample},'' {\em Astrophys.J.}, vol.~746, p.~85, 2012.

\bibitem{Sullivan:2011kv}
M.~Sullivan {\em et~al.}, ``{SNLS3: Constraints on Dark Energy Combining the
  Supernova Legacy Survey Three Year Data with Other Probes},'' {\em
  Astrophys.J.}, vol.~737, p.~102, 2011.

\bibitem{Tegmark:1998yy}
M.~Tegmark, D.~J. Eisenstein, W.~Hu, and R.~G. Kron, ``{Cosmic complementarity:
  Probing the acceleration of the universe},'' 1998.

\bibitem{Kogut:1993ag}
A.~Kogut, C.~Lineweaver, G.~F. Smoot, C.~Bennett, A.~Banday, {\em et~al.},
  ``{Dipole anisotropy in the COBE DMR first year sky maps},'' {\em
  Astrophys.J.}, vol.~419, p.~1, 1993.

\bibitem{Kashlinsky:2008ut}
A.~Kashlinsky, F.~Atrio-Barandela, D.~Kocevski, and H.~Ebeling, ``{A
  measurement of large-scale peculiar velocities of clusters of galaxies:
  results and cosmological implications},'' {\em Astrophys.J.}, vol.~686,
  pp.~L49--L52, 2009.

\bibitem{Kashlinsky:2009dw}
A.~Kashlinsky, F.~Atrio-Barandela, H.~Ebeling, A.~Edge, and D.~Kocevski, ``{A
  new measurement of the bulk flow of X-ray luminous clusters of galaxies},''
  {\em Astrophys.J.}, vol.~712, pp.~L81--L85, 2010.

\bibitem{Kashlinsky:2008us}
A.~Kashlinsky, F.~Atrio-Barandela, D.~Kocevski, and H.~Ebeling, ``{A
  measurement of large-scale peculiar velocities of clusters of galaxies:
  technical details},'' {\em Astrophys.J.}, vol.~691, pp.~1479--1493, 2009.

\bibitem{Kashlinsky:2010ur}
A.~Kashlinsky, F.~Atrio-Barandela, and H.~Ebeling, ``{Measuring the dark flow
  with public X-ray cluster data},'' {\em Astrophys.J.}, vol.~732, p.~1, 2011.

\bibitem{Turner:1991dn}
M.~S. Turner, ``{A Tilted Universe (and Other Remnants of the Preinflationary
  Universe)},'' {\em Phys.Rev.}, vol.~D44, pp.~3737--3748, 1991.

\bibitem{Afshordi:2008rd}
N.~Afshordi, G.~Geshnizjani, and J.~Khoury, ``{Do observations offer evidence
  for cosmological-scale extra dimensions?},'' {\em JCAP}, vol.~0908, p.~030,
  2009.

\bibitem{Campanelli:2009tk}
L.~Campanelli, ``{A Model of Universe Anisotropization},'' {\em Phys.Rev.},
  vol.~D80, p.~063006, 2009.

\bibitem{LyndenBell:1988qs}
D.~Lynden-Bell, S.~Faber, D.~Burstein, R.~L. Davies, A.~Dressler, {\em et~al.},
  ``{Spectroscopy and photometry of elliptical galaxies. V - Galaxy streaming
  toward the new supergalactic center},'' {\em Astrophys.J.}, vol.~326, p.~19,
  1988.

\bibitem{Scaramella89}
R.~S. G. B.-P. G. C. G. V.~G. Zamorani {\em Nature}, vol.~338, p.~562, 1989.

\bibitem{Watkins:2008hf}
R.~Watkins, H.~A. Feldman, and M.~J. Hudson, ``{Consistently Large Cosmic Flows
  on Scales of 100 Mpc/h: a Challenge for the Standard LCDM Cosmology},'' {\em
  Mon.Not.Roy.Astron.Soc.}, vol.~392, pp.~743--756, 2009.

\bibitem{Turnbull:2011ty}
S.~J. Turnbull, M.~J. Hudson, H.~A. Feldman, M.~Hicken, R.~P. Kirshner, {\em
  et~al.}, ``{Cosmic flows in the nearby universe from Type Ia Supernovae},''
  {\em Mon.Not.Roy.Astron.Soc.}, vol.~420, pp.~447--454, 2012.

\bibitem{Colin:2010ds}
J.~Colin, R.~Mohayaee, S.~Sarkar, and A.~Shafieloo, ``{Probing the anisotropic
  local universe and beyond with SNe Ia data},'' {\em Mon.Not.Roy.Astron.Soc.},
  vol.~414, pp.~264--271, 2011.

\bibitem{Nusser:2011sd}
A.~Nusser, E.~Branchini, and M.~Davis, ``{Bulk flows from galaxy luminosities:
  application to 2MASS redshift survey and forecast for next-generation
  datasets},'' {\em Astrophys.J.}, vol.~735, p.~77, 2011.

\bibitem{Nusser:2011tu}
A.~Nusser and M.~Davis, ``{The cosmological bulk flow: consistency with
  $\Lambda$CDM and $z\approx 0$ constraints on $\sigma_8$ and $\gamma$},'' {\em
  Astrophys.J.}, vol.~736, p.~93, 2011.

\bibitem{Branchini:2012rb}
E.~Branchini, M.~Davis, and A.~Nusser, ``{The velocity field of 2MRS Ks=11.75
  galaxies: constraints on beta and bulk flow from the luminosity function},''
  {\em Mon.Not.Roy.Astron.Soc.}, vol.~424, pp.~472--481, 2012.

\bibitem{Ma:2012tt}
Y.-Z. Ma and D.~Scott, ``{Cosmic bulk flows on 50 ${h}^{-1}$Mpc scales: A
  Bayesian hyper-parameter method and multi-shells likelihood analysis},'' {\em
  Mon.Not.Roy.Astron.Soc.}, vol.~428, p.~2017, 2013.

\bibitem{Keisler:2009nw}
R.~Keisler, ``{The Statistical Significance of the 'Dark Flow'},'' {\em
  Astrophys.J.}, vol.~707, pp.~L42--L44, 2009.

\bibitem{Osborne:2010mf}
S.~Osborne, D.~Mak, S.~Church, and E.~Pierpaoli, ``{Measuring the Galaxy
  Cluster Bulk Flow from WMAP data},'' {\em Astrophys.J.}, vol.~737, p.~98,
  2011.

\bibitem{Mody:2012rh}
K.~Mody and A.~Hajian, ``{One Thousand and One Clusters: Measuring the Bulk
  Flow with the Planck ESZ and X-Ray Selected Galaxy Cluster Catalogs},'' {\em
  Astrophys.J.}, vol.~758, p.~4, 2012.

\bibitem{Lavaux:2012jb}
G.~Lavaux, N.~Afshordi, and M.~J. Hudson, ``{First measurement of the bulk flow
  of nearby galaxies using the cosmic microwave background},'' {\em
  Mon.Not.Roy.Astron.Soc.}, vol.~430, pp.~1617--1635, 2013.

\bibitem{Abate:2012za}
A.~Abate {\em et~al.}, ``{Large Synoptic Survey Telescope: Dark Energy Science
  Collaboration},'' 2012.

\bibitem{Johnson:2014kaa}
A.~Johnson, C.~Blake, J.~Koda, Y.-Z. Ma, M.~Colless, {\em et~al.}, ``{The 6dF
  Galaxy Velocity Survey: Cosmological constraints from the velocity power
  spectrum},'' {\em Mon.Not.Roy.Astron.Soc.}, vol.~444, p.~3926, 2014.

\bibitem{Feindt:2013pma}
U.~Feindt, M.~Kerschhaggl, M.~Kowalski, G.~Aldering, P.~Antilogus, {\em
  et~al.}, ``{Measuring cosmic bulk flows with Type Ia Supernovae from the
  Nearby Supernova Factory},'' {\em Astron.Astrophys.}, 2013.

\bibitem{Appleby:2013ida}
S.~Appleby and A.~Shafieloo, ``{Testing local anisotropy using the method of
  smoothed residuals I — methodology},'' {\em JCAP}, vol.~1403, p.~007, 2014.

\bibitem{Kolatt:2000yg}
T.~S. Kolatt and O.~Lahav, ``{Constraints on cosmological anisotropy out to z=1
  from supernovae ia},'' {\em Mon.Not.Roy.Astron.Soc.}, vol.~323, p.~859, 2001.

\bibitem{Bonvin:2006en}
C.~Bonvin, R.~Durrer, and M.~Kunz, ``{The dipole of the luminosity distance: a
  direct measure of h(z)},'' {\em Phys.Rev.Lett.}, vol.~96, p.~191302, 2006.

\bibitem{Gordon:2007zw}
C.~Gordon, K.~Land, and A.~Slosar, ``{Cosmological Constraints from Type Ia
  Supernovae Peculiar Velocity Measurements},'' {\em Phys.Rev.Lett.}, vol.~99,
  p.~081301, 2007.

\bibitem{Schwarz:2007wf}
D.~J. Schwarz and B.~Weinhorst, ``{(An)isotropy of the Hubble diagram:
  Comparing hemispheres},'' {\em Astron.Astrophys.}, vol.~474, pp.~717--729,
  2007.

\bibitem{Gupta:2007pb}
S.~Gupta, T.~D. Saini, and T.~Laskar, ``{Direction Dependent Non-Gaussianity in
  High-z Supernova Data},'' {\em Mon.Not.Roy.Astron.Soc.}, vol.~388,
  pp.~242--246, 2008.

\bibitem{Appleby:2012as}
S.~A. Appleby and E.~V. Linder, ``{Probing Dark Energy Anisotropy},'' {\em
  Phys.Rev.}, vol.~D87, p.~023532, 2013.

\bibitem{Cooray:2008qn}
A.~R. Cooray, D.~E. Holz, and R.~Caldwell, ``{Measuring dark energy spatial
  inhomogeneity with supernova data},'' {\em JCAP}, vol.~1011, p.~015, 2010.

\bibitem{Gupta:2010jp}
S.~Gupta and T.~D. Saini, ``{Direction Dependence in Supernova Data:
  Constraining Isotropy},'' 2010.

\bibitem{Cooke:2009ws}
R.~Cooke and D.~Lynden-Bell, ``{Does the Universe Accelerate Equally in all
  Directions?},'' {\em Mon.Not.Roy.Astron.Soc.}, vol.~401, pp.~1409--1414,
  2010.

\bibitem{Koivisto:2008ig}
T.~Koivisto and D.~F. Mota, ``{Anisotropic Dark Energy: Dynamics of Background
  and Perturbations},'' {\em JCAP}, vol.~0806, p.~018, 2008.

\bibitem{Koivisto:2010dr}
T.~S. Koivisto, D.~F. Mota, M.~Quartin, and T.~G. Zlosnik, ``{On the
  Possibility of Anisotropic Curvature in Cosmology},'' {\em Phys.Rev.},
  vol.~D83, p.~023509, 2011.

\bibitem{Campanelli:2007qn}
L.~Campanelli, P.~Cea, and L.~Tedesco, ``{Cosmic Microwave Background
  Quadrupole and Ellipsoidal Universe},'' {\em Phys.Rev.}, vol.~D76, p.~063007,
  2007.

\bibitem{Campanelli:2006vb}
L.~Campanelli, P.~Cea, and L.~Tedesco, ``{Ellipsoidal Universe Can Solve The
  CMB Quadrupole Problem},'' {\em Phys.Rev.Lett.}, vol.~97, p.~131302, 2006.

\bibitem{Appleby:2009za}
S.~Appleby, R.~Battye, and A.~Moss, ``{Constraints on the anisotropy of dark
  energy},'' {\em Phys.Rev.}, vol.~D81, p.~081301, 2010.

\bibitem{Antoniou:2010gw}
I.~Antoniou and L.~Perivolaropoulos, ``{Searching for a Cosmological Preferred
  Axis: Union2 Data Analysis and Comparison with Other Probes},'' {\em JCAP},
  vol.~1012, p.~012, 2010.

\bibitem{Blomqvist:2008ud}
M.~Blomqvist, E.~Mortsell, and S.~Nobili, ``{Probing Dark Energy
  Inhomogeneities with Supernovae},'' {\em JCAP}, vol.~0806, p.~027, 2008.

\bibitem{Blomqvist:2010ky}
M.~Blomqvist, J.~Enander, and E.~Mortsell, ``{Constraining dark energy
  fluctuations with supernova correlations},'' {\em JCAP}, vol.~1010, p.~018,
  2010.

\bibitem{Tsagas:2009nh}
C.~G. Tsagas, ``{Large-scale peculiar motions and cosmic acceleration},'' {\em
  Mon.Not.Roy.Astron.Soc.}, vol.~405, p.~503, 2010.

\bibitem{Appleby:2014lra}
S.~Appleby and A.~Shafieloo, ``{Testing Isotropy in the Local Universe},'' {\em
  JCAP}, vol.~1410, p.~070, 2014.

\bibitem{Hui:2005nm}
L.~Hui and P.~B. Greene, ``{Correlated Fluctuations in Luminosity Distance and
  the (Surprising) Importance of Peculiar Motion in Supernova Surveys},'' {\em
  Phys.Rev.}, vol.~D73, p.~123526, 2006.

\bibitem{Davis:2010jq}
T.~M. Davis, L.~Hui, J.~A. Frieman, T.~Haugbolle, R.~Kessler, {\em et~al.},
  ``{The effect of peculiar velocities on supernova cosmology},'' {\em
  Astrophys.J.}, vol.~741, p.~67, 2011.

\bibitem{Kowalski:2008ez}
M.~Kowalski {\em et~al.}, ``{Improved Cosmological Constraints from New, Old
  and Combined Supernova Datasets},'' {\em Astrophys.J.}, vol.~686,
  pp.~749--778, 2008.

\bibitem{Perivolaropoulos:2008yc}
L.~Perivolaropoulos and A.~Shafieloo, ``{Bright High z SnIa: A Challenge for
  Lambda CDM?},'' {\em Phys.Rev.}, vol.~D79, p.~123502, 2009.

\bibitem{Strauss:1995fz}
M.~A. Strauss and J.~A. Willick, ``{The Density and peculiar velocity fields of
  nearby galaxies},'' {\em Phys.Rept.}, vol.~261, pp.~271--431, 1995.

\bibitem{Ma:2010ps}
Y.-Z. Ma, C.~Gordon, and H.~A. Feldman, ``{The peculiar velocity field:
  constraining the tilt of the Universe},'' {\em Phys.Rev.}, vol.~D83,
  p.~103002, 2011.

\bibitem{Zehavi:1998gz}
I.~Zehavi, A.~G. Riess, R.~P. Kirshner, and A.~Dekel, ``{A Local hubble bubble
  from SNe Ia?},'' {\em Astrophys.J.}, vol.~503, p.~483, 1998.

\bibitem{Conley:2007ng}
A.~J. Conley {\em et~al.}, ``{Is there Evidence for a Hubble bubble? The Nature
  of Type Ia Supernova Colors and Dust in External Galaxies},'' {\em
  Astrophys.J.}, vol.~664, pp.~L13--L16, 2007.

\bibitem{Jha:2006fm}
S.~Jha, A.~G. Riess, and R.~P. Kirshner, ``{Improved Distances to Type Ia
  Supernovae with Multicolor Light Curve Shapes: MLCS2k2},'' {\em
  Astrophys.J.}, vol.~659, pp.~122--148, 2007.

\bibitem{Hamuy:1996su}
M.~Hamuy, M.~Phillips, N.~B. Suntzeff, R.~A. Schommer, and J.~Maza, ``{BVRI
  Light Curves for 29 Type Ia Supernovae},'' {\em Astron.J.}, vol.~112,
  pp.~2408--2437, 1996.

\bibitem{Jha:2005jg}
S.~Jha, R.~P. Kirshner, P.~Challis, P.~M. Garnavich, T.~Matheson, {\em et~al.},
  ``{Ubvri light curves of 44 type ia supernovae},'' {\em Astron.J.}, vol.~131,
  pp.~527--554, 2006.

\bibitem{Contreras:2009nt}
C.~Contreras, M.~Hamuy, M.~Phillips, G.~Folatelli, N.~B. Suntzeff, {\em
  et~al.}, ``{The Carnegie Supernova Project: First Photometry Data Release of
  Low-Redshift Type Ia Supernovae},'' {\em Astron.J.}, vol.~139, pp.~519--539,
  2010.

\bibitem{Blondin:2012ha}
S.~Blondin, T.~Matheson, R.~Kirshner, K.~Mandel, P.~Berlind, {\em et~al.},
  ``{The Spectroscopic Diversity of Type Ia Supernovae},'' {\em Astron.J.},
  vol.~143, p.~126, 2012.

\bibitem{Gibelyou:2012ri}
C.~Gibelyou and D.~Huterer, ``{Dipoles in the Sky},'' {\em
  Mon.Not.Roy.Astron.Soc.}, vol.~427, pp.~1994--2021, 2012.

\end{thebibliography}
\bibliographystyle{ieeetr}

\end{document}